\documentclass{emulateapj}

\usepackage{graphicx}

\usepackage{mathrsfs, amsmath}

\newcommand{\su}{S09}

\slugcomment{accepted to The Astrophysical Journal on April 4, 2016}
\shorttitle{Imaging HR 8799 with STIS}
\shortauthors{Gerard, B. et al.}

\begin{document}
\title{Searching for the HR 8799 Debris Disk with \textit{HST}/STIS}
\author{Gerard, B.\altaffilmark{1,2}}
\email{bgerard@uvic.ca}
\author{Lawler, S.\altaffilmark{2}, Marois, C. \altaffilmark{2,1}, Tannock, M.\altaffilmark{1}, Matthews, B.\altaffilmark{2,1}, Venn, K.\altaffilmark{1}}
\altaffiltext{1}{University of Victoria, Department of Physics and Astronomy, 3800 Finnerty Rd, Victoria, V8P 5C2, Canada}
\altaffiltext{2}{National Research Council of Canada, Astronomy \& Astrophysics Program, 5071 West Saanich Rd, Victoria, V9E 2E7, Canada}
\begin{abstract}
We present a new algorithm for space telescope high contrast imaging of close-to-face-on planetary disks called Optimized Spatially Filtered  (OSFi) normalization. This algorithm is used on HR 8799 \textit{Hubble Space Telescope (HST)} coronagraphic archival data, showing an over-luminosity after reference star point spread function (PSF) subtraction that may be from the inner disk and/or planetesimal belt components of this system. The PSF-subtracted radial profiles in two separate epochs from 2011 and 2012 are consistent with one another, and self-subtraction shows no residual in both epochs. We explore a number of possible false-positive scenarios that could explain this residual flux, including telescope breathing, spectral differences between HR 8799 and the reference star, imaging of the known warm inner disk component, OSFi algorithm throughput and consistency with the standard spider normalization \textit{HST} PSF subtraction technique, and coronagraph misalignment from pointing accuracy. In comparison to another similar STIS dataset, we find that the over-luminosity is likely a result of telescope breathing and spectral difference between HR 8799 and the reference star. Thus, assuming a non-detection, we derive upper limits on the HR 8799 dust belt mass in small grains. In this scenario, we find that the flux of these micron-sized dust grains leaving the system due to radiation pressure is small enough to be consistent with measurements of other debris disk halos.
\end{abstract}
\section{Introduction}
\label{intro}
The era of direct imaging of extrasolar planets is upon us. High contrast images of HR 8799 have revealed the presence of four planets orbiting their host star \citep{marois08, marois10}, and more recent integral field spectrographs have provided spectra of their atmospheres \citep{bowler, barman, gpi8799}. In contrast to radial velocity or transit exoplanet detection methods, these giant planets at large separations trace a range of parameter space  often closer to our own Solar System, and so understanding the physical and chemical properties of this and other similar systems is crucial to better understanding the process of Solar System and planet formation.

One key step in better understanding the planet formation process is the formation and stability of protoplanetary disks and their remnant debris disks. The HR 8799 debris disk was first measured by \cite{first8799} from an unresolved spectral energy distribution (SED) infrared excess at 60 $\mu$m with the Infrared Astronomical Telescope (\textit{IRAS}) point source catalogue. The debris disk was later slightly spatially resolved by \citet{su} (hereafter \su) also with an SED infrared excess. \su\; used the \textit{Spitzer Space Telescope} at 24, 70, and 160 $\mu$m to measure one component inside the known orbiting planets \cite[the planets are between 15 and 68 AU;][]{marois08, marois10} at $\sim6-15$ AU---the inner disk---and another two components outside the planets at $\sim$90-300 AU---the planetesimal belt---and $\sim$300-1000 AU---the halo. \cite{Brenda} have also spatially resolved the HR 8799 planetesimal belt and halo at 70, 100, and 160 $\mu$m and marginally resolved at 250, 350, and 500 $\mu$m with the \textit{Herschel Space Telescope}, also measuring a similar three component debris disk from image and surface brightness profile modelling. \cite{Brenda} also measure a disk inclination of $26\pm3^{\circ}$, close to face-on orientation.

However, none of the HR 8799 debris disk components have been imaged in optical scattered light due to the close-to-face-on nature of the disk. Angular Differential Imaging (ADI) point spread function (PSF) subtraction \citep{adi}, a standard optical/near infrared high contrast imaging technique used to construct a reference stellar/instrumental PSF from a sequence of science images, is not optimized for close-to-face-on planetary disks and causes self-subtraction effects. Additionally, a low spatial frequency close-to-face-on disk is hard to distinguish from focus and pointing errors, which at minimum need to be kept stable at the contrast needed for a detection. High contrast close-to-face-on disk imaging with ground based telescopes is generally avoided due to both limited stability from atmospheric turbulence, even with adaptive optics, and self-subtraction effects. Observing a separate reference star of matching spectral type in the same sequence as the target star using ground based telescopes is also generally avoided for close-to-face-on disks due to limited stability. Imaging close-to-face-on disks from space is still difficult but more feasible in both highly correlated stability and by observing a separate star as the reference PSF to prevent self-subtraction effects.More recent robust least-squares-based PSF subtraction analysis using Hubble Space Telescope (\textit{HST}) data \citep[e.g.,][]{klip,rafael,currie,alice,alice1,alice2} have made significant improvements over classical PSF subtraction techniques (e.g., see \S\ref{spiders}).

The optical scattering properties of small grains in the HR 8799 debris disk (i.e., the contrast needed for a \textit{HST} detection) is unknown. The \textit{HST} Space Telescope Imaging Spectrograph (STIS) imaging CCD can be used with a Lyot coronagraph to reach contrasts similar to ground based extreme adaptive optics instruments \citep[$\sim10^{-6}$, e.g., ][]{schneider}. The relatively wide 52 $\times$ 52 arcsecond STIS field of view provides an optimal opportunity to image the HR 8799 planetesimal belt in optical wavelengths.

In this paper we present a new PSF subtraction algorithm, called Optimized Spatially Filtered  (OSFi) normalization, that was developed to image the HR 8799 planetesimal belt with STIS. In \S\ref{data} we describe the STIS data. In \S\ref{PSF_subtraction} we describe our new OSFi normalization algorithm. In \S\ref{usable_data} we present the results of OSFi normalization applied to the HR 8799 STIS data. In \S\ref{heuristics} we analyze a number of false-positive detection heuristics based on our OSFi algorithm results, ultimately indicating that we are seeing a non-detection. In \S\ref{science} we present an upper limit analysis to constrain the mass contained in the sub-micron grain planetesimal belt, and in \S\ref{conclusion} we conclude and summarize our results. Throughout this paper we use the following measurements for HR 8799: distance $d_\star= 39.9$ pc, radius $R_\star=1.34 R_\odot$, effective temperature $T_\star$ = 7250 K, stellar luminosity $L_\star=4.9\; L_\odot,\; \text{and stellar mass } M_\star=1.5\; M_\odot$ \cite[]{sadakane, gray, su, Brenda}. The HR 8799 spectral type varies in the literature between A5V and F0V with characteristics of an anomalous $\lambda$ Boo star \cite[]{gray}. The STIS plate scale is 0.05078 arcseconds per pixel \cite[]{stis_handbook}.
\section{STIS HR 8799 Data}
\label{data}
We obtained public archival STIS data of HR 8799 \citep{archive} from the Mukilski Archive for Space Telescopes (MAST). STIS coronographic observations of HR 8799 are present from three epochs: 2011 November 2 (hereafter epoch 1), 2011 November 12, and 2012 October 4 (hereafter epoch 2). Epochs 1 and 2 each contain three 2300s exposures of HR 8799, one exposure per \textit{HST} orbit. The three successive images in epoch 1 and three successive images in epoch 2 are hereafter referred to as science images 0, 1, 2, 3, 4, and 5, respectively. One 2300s exposure of HIP 117990, a separate reference star of similar color to HR 8799, is present in the final orbit of epoch 1 and 2 (hereafter referred to as reference images 1 and 2, respectively). Data from the 2011 November 12 epoch contained nine 120 s and three 240 s science exposures and one 2226 s reference exposure and is therefore unusable due to the different noise floors between reference and science images. All of the HR 8799 images within a given epoch are taken at different telescope roll angles. We use the dark-subtracted, flat-fielded, cosmic ray-rejected {\tt{.crj}} images \cite[]{stis_handbook}.
\section{OSFi normalization: A New PSF subtraction algorithm}
\label{PSF_subtraction}
\subsection{Image Registration}
\label{image_registration}

Before we can implement PSF subtraction between a science and reference image (hereafter called reference-subtraction) or two different science images (hereafter called self-subtraction), all images must be registered to a common center with sub-pixel accuracy. To do this, after normalizing the flux in each image to 1/(exposure time), we place an aperture mask around the four diffraction ``spiders'' in the regions unaffected by the coronagraph focal plane mask (FPM) and set the remainder of the image to zero. We then run upsampled cross-correlation on the masked spider image with the same image rotated 180 degrees about its center using an image registration Python package\footnote{https://github.com/keflavich/image\_registration} to determine the image center to sub-pixel accuracy and then shift the unmasked images to a common center. We used cubic spline interpolation for sub-pixel shifting to avoid ringing effects from a Fourier-based shifting algorithm.
\subsection{OSFi PSF subtraction}
\label{psf_subtraction}

A classical least-squares locally optimized combination of images (LOCI) PSF subtraction \citep{loci} is not optimal to minimize the noise by constructing a reference image from the science images for a dataset with only two images per epoch\footnote{Science images 0 and 3 are unusable due to decreased pointing stability (\S\ref{usable_data}).}. Additionally, as mentioned in \S\ref{intro}, constructing the reference image from the science images with ADI will cause self-subtraction of the close-to-face-on HR 8799 debris disk. 

We also considered the option of using reference differential imaging (RDI) through the Archival Legacy Investigation of Circumstellar Environments (ALICE) pipeline \citep{alice}, where a large archive of HST reference images is used in a principal component analysis-based least-squares calculation \citep{klip}. Although RDI removes the risk of self-subtraction for disks, we concluded that this approach is not optimal for HR 8799. Because we are looking for a close-to-face-on disk with a smooth radial profile \citep{Brenda}, using additional reference images from different epochs may have (1) been observed at a slightly different relative focus from the HR 8799 observations, and (2) a different stellar spectral type compared to HR 8779, which we will see in \S\ref{heuristics} can cause a similar radial profile that can be confused with a disk.

Instead, normalizing the reference PSF subtraction (i.e., using the single separately acquired reference image in the same sequence as HR 8799) to the high spatial frequency noise can be thought of as ``minimizing'' the noise contaminating the expected low spatial frequency, close-to-face-on disk. This is, in essence, a least-squares using a locally optimized region for face-on disks that is designed to work on small datasets. This is exactly the rationale behind an Optimized Spatially Filtered (OSFi) PSF subtraction: we subtract the high spatial frequency noise from the science image, leaving the residual low spatial frequency astrophysical disk. This is done by normalizing the subtraction of a separately acquired reference image (im$_\text{ref}$) from the science image (im$_\text{sci}$) to the high spatial frequency noise in each PSF:
\begin{equation}
\text{im}_{\text{OSFi}}=\text{im}_\text{sci}-\text{im}_\text{ref}\left(\frac{\sigma\left[ \mathscr{F}_H \left( \text{im}_\text{sci}\right) \right]}{\sigma\left[ \mathscr{F}_H \left(\text{im}_\text{ref} \right) \right]}\right),
\label{eq: algo}
\end{equation}
where $\sigma$ is the robust standard deviation operator \citep{robust} in a user-defined optimization region sampling the PSF, $\mathscr{F}_H$ is the image high-pass filter operator, and $\text{im}_{\text{OSFi}}$ is the residual PSF-subtracted image. Although there are many methods of high-pass filtering in image processing (e.g., Fourier filtering, convolution kernels, etc.), our initial approach was to first create a low-pass filtered image, $\mathscr{F}_L (\text{im})$, using a basic 11 pixel-wide median boxcar filter and then create the high-pass image, $\mathscr{F}_H (\text{im})$, via:
\begin{equation}
\mathscr{F}_H (\text{im})=\text{im}-\mathscr{F}_L (\text{im})
\label{eq: im_high}
\end{equation}

\begin{figure}[!h]
\centering
\plotone{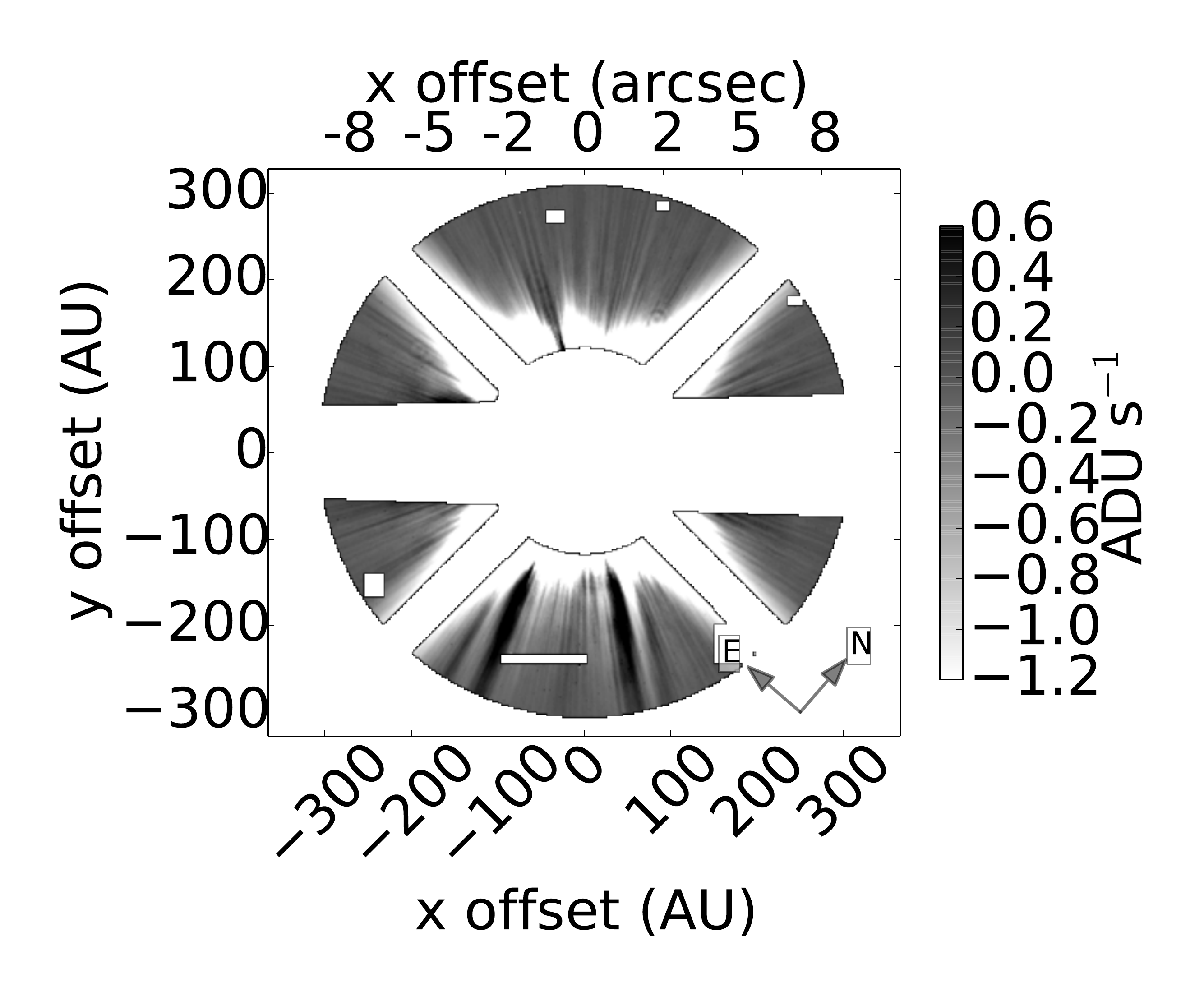}
\caption{An example of a typical OSFi optimization region (here showing science image 1) after both masking the coronagraph, diffraction spiders, and bad pixels and high-pass filtering the image. The coordinate grid shows x and y offset from the registered image center in both arcseconds and AU.}
\label{fig: optimization_region}
\end{figure}

In \S\ref{heuristics} we further discuss the dependency of different high-pass filtering techniques on our OSFi algorithm throughput. We found that an 11 pixel standard deviation Gaussian convolution kernel used to construct  $\mathscr{F}_L (\text{im})$, and by equation \ref{eq: im_high} $\mathscr{F}_H (\text{im})$, has the best throughput performance in equation \ref{eq: algo}, and so we use this technique for the rest of our analysis (see \S\ref{algo}.1). We chose our optimization region as an annulus between 3.5 arcseconds (140 AU; the STIS region outside which there are no saturated pixels leaking from behind the FPM, the effective inner working angle) and 7.5 arcseconds (300 AU) in agreement with the measured $\sim$90-300 AU planetesimal belt in \su. To create the optimization region, the image is first median filtered using a 3 by 3 pixel boxcar to remove individual bad pixels. Masks are then also placed over the spiders, coronagraph wedge, and any additional remaining bad pixels/regions before high-pass filtering with an 11 pixel standard deviation Gaussian kernel. An example high spatial frequency optimization region is shown in Figure \ref{fig: optimization_region}. 

Finally, multiple PSF-subtracted images taken at different telescope roll angles within a single epoch can be combined to additionally reduce the residual quasi-static high spatial frequency noise by rotating each PSF-subtracted image to north up and median combining all de-rotated images. This does not cause ADI self-subtraction as discussed in \S\ref{intro} because the reference image is a separate, stellar PSF with no disk.

Self-subtraction using the OSFi algorithm should yield a featureless noise map because there is a close-to-face-on disk in both images. This is essentially a measure of contrast if both science images have the same diffuse close-to-face-on halo features that are independent of telescope roll angle and the low spatial frequencies of the two PSFs are normalized. As we will see in \S \ref{usable_data}, the latter is not the case if the observing sequence experiences a focus evolution due to telescope breathing.

On a similar note, we further emphasize that our OSFi algorithm is designed for small, correlated datasets. OSFi RDI would not work well to image a diffuse face-on halo because we are only subtracting the high spatial frequencies of the PSF, assuming that the low spatial frequencies are normalized to the same coefficient. This is not a fair assumption if the observing sequence experiences a focus evolution (\S\ref{usable_data},\ref{heuristics}) and/or a difference between the reference and target star spectral type (\S\ref{heuristics}). These are low spatial frequency effects that are not seen by the OSFi algorithm due to high pass filtering, and so including a larger diversity of datasets with a different spectral type and focus evolution effects would ultimately limit OSFi sensitivity to detect a face-on, diffuse disk rather than improve it.
\section{Results}
\label{usable_data}
\begin{figure*}[!ht]
\centering
\plottwo{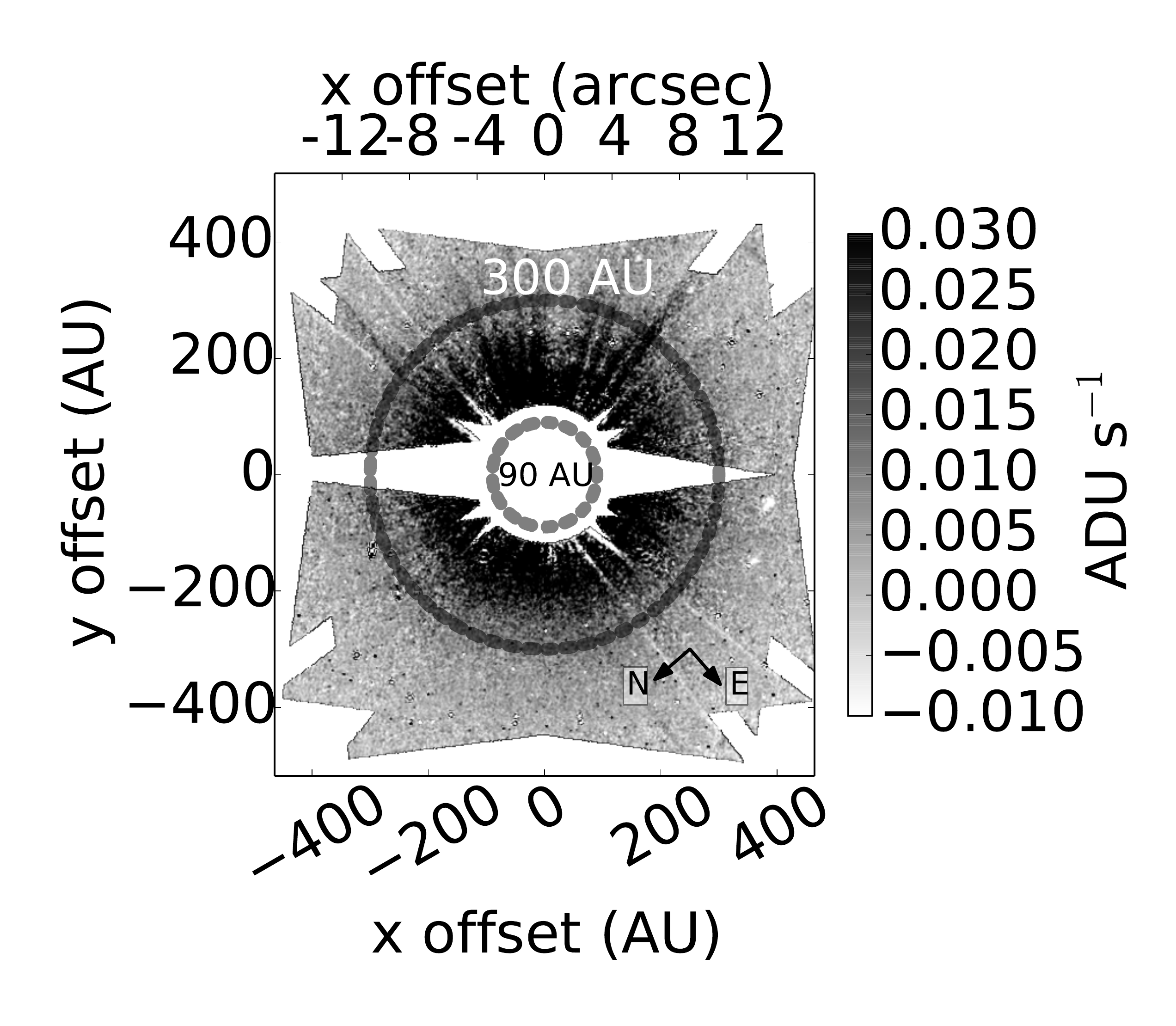}{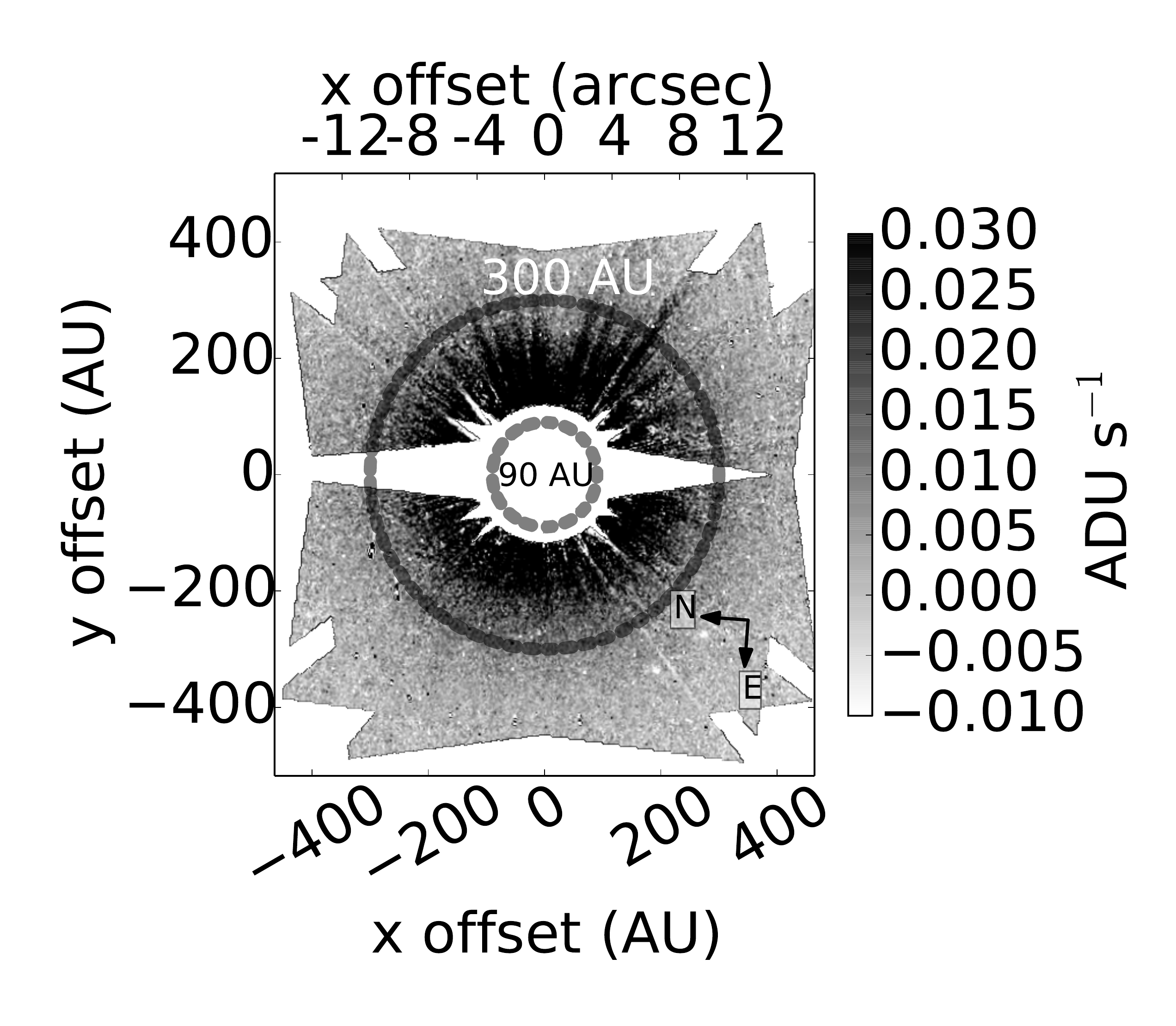}
\caption{Reference-subtracted OSFi-normalized images from 2011, November 2 (left; epoch 1), and 2012, October 4 (right; epoch 2), using the separately acquired reference images of HIP 117990 during the same observing sequence to subtract from the HR 8799 images. The observed residual over-luminosity is consistent between both epochs. All following PSF-subtracted images are derotated to the same matching pupil orientation for comparison and show the measured \su\; 90--300 AU planetesimal belt region outlined by dashed circles.}
\label{fig: ref_subtract}
\end{figure*}
\begin{figure*}[!hb]
\centering
\plottwo{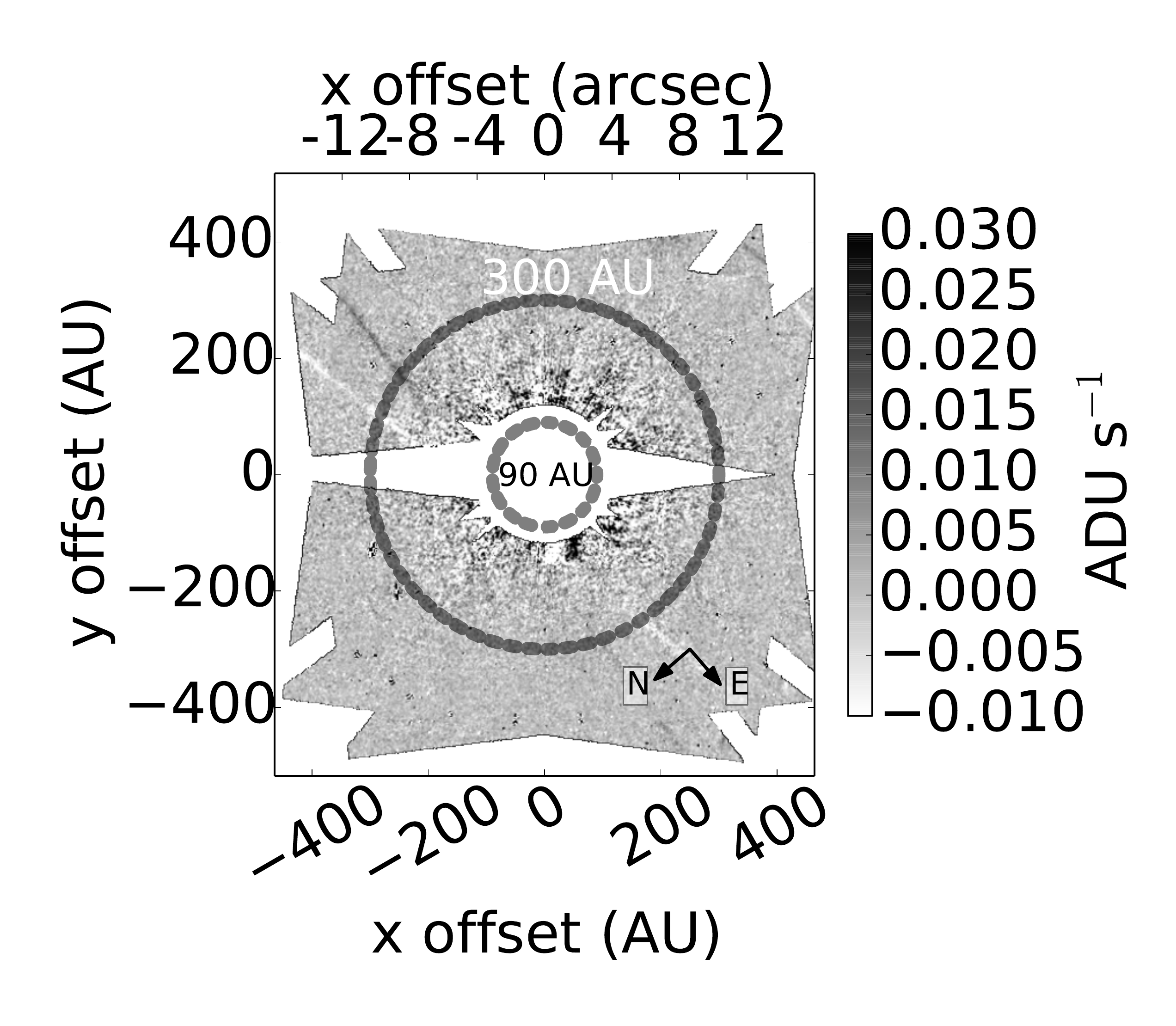}{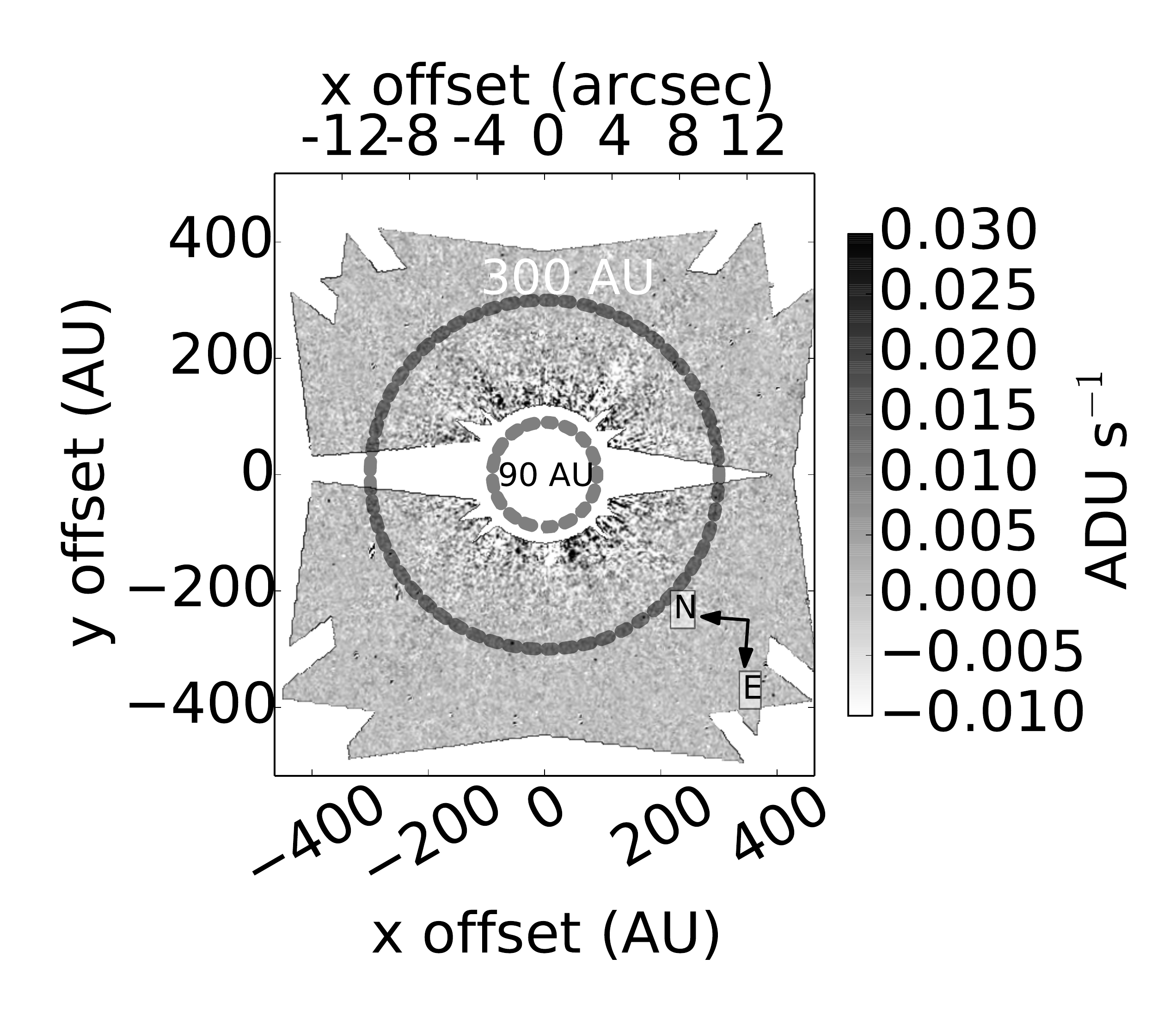}
\caption{The epoch 1 (left) and epoch 2 (right) self-subtracted, OSFi-normalized images. Here, the reference image is a science image at the same epoch but different roll angle, and so the displayed featureless residuals from OSFi PSF subtraction show that the same low spatial frequencies in both science images are independent of telescope roll angle (i.e., there is no detected focus evolution).}
\label{fig: self_subtract}
\end{figure*}
\begin{figure}[!h]
\centering
\plotone{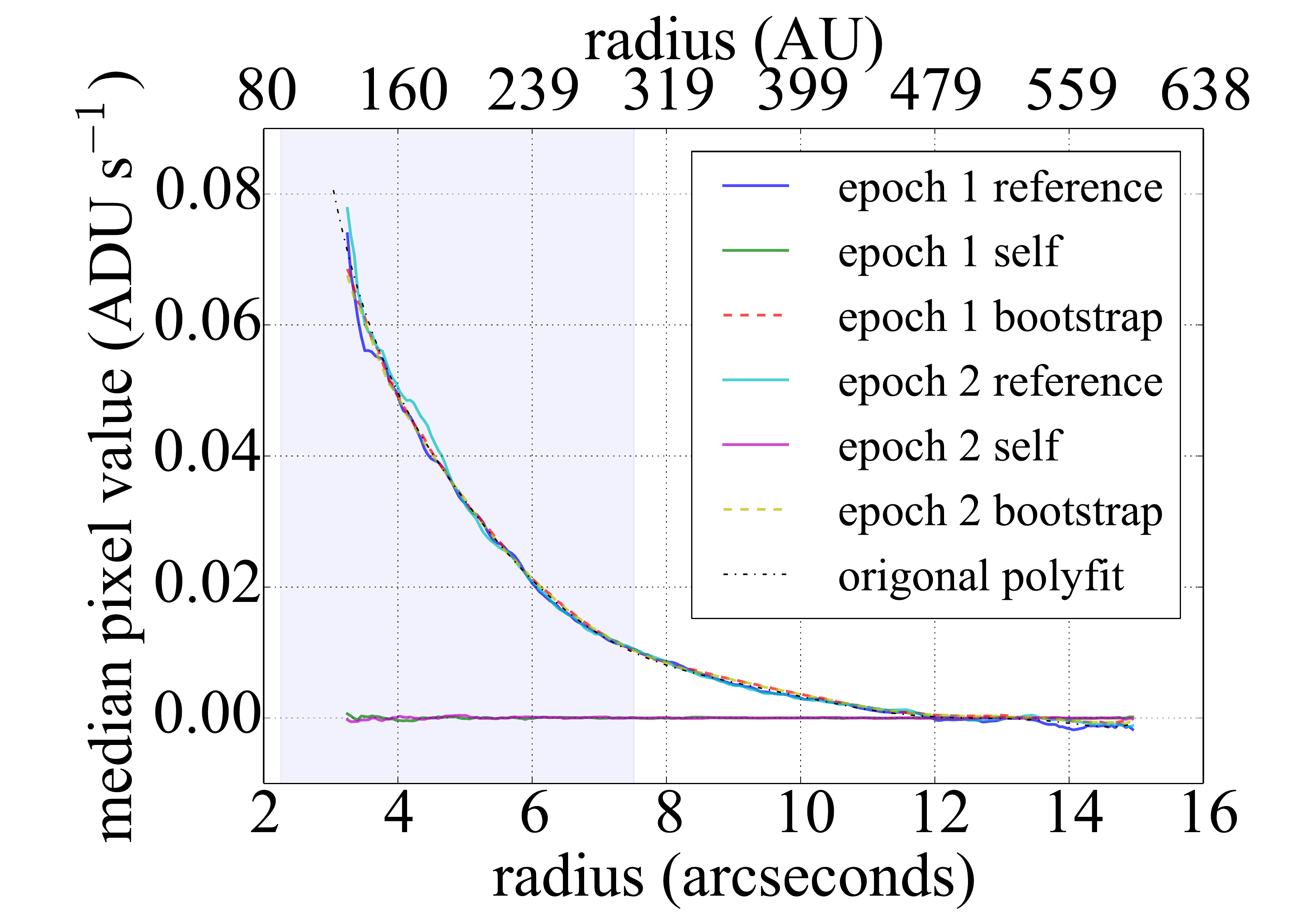}
\caption{Radial profiles for OSFi reference- and self-subtracted images (\S \ref{usable_data}), bootstrapped images (\S \ref{boot}), and a fitted polynomial profile to reference-subtracted images (\S \ref{boot}). The reference subtracted images use HIP 117990 as the reference star, whereas the self-subtracted images use HR 8799 at a different telescope roll angle as the reference star. The bootstrapped images use HR 8799 at a different telescope roll angle as a reference image, but the raw HR 8799 target image includes an added polynomial radial profile to measure our OSFi algorithm throughput. The shaded 90--300 AU region represents the measured \su\; planetesimal belt.}
\label{fig: OSFi_rad}
\end{figure}
OSFi reference-subtraction and self-subtraction for both epochs are shown in Figures \ref{fig: ref_subtract} and \ref{fig: self_subtract}, respectively. The images have been cropped to the central $\sim$30 arcsecond region of interest, derotated to the match pupil orientation in each respective epoch, and show the measured \su\; 90--300 AU planetesimal belt region. Reference-subtraction in shows a diffuse, close-to-face-on halo, whereas self-subtraction shows hardly any residual flux. The radial profiles for both epochs of self- and reference-subtraction are shown in Figure \ref{fig: OSFi_rad} as the solid lines, illustrating that reference-subtraction is consistent between both epochs and confirming that self-subtraction yields no residual.
\begin{figure*}[!h]
  \begin{minipage}[b]{0.5\linewidth}
    \centering
    \includegraphics[width=0.65\linewidth]{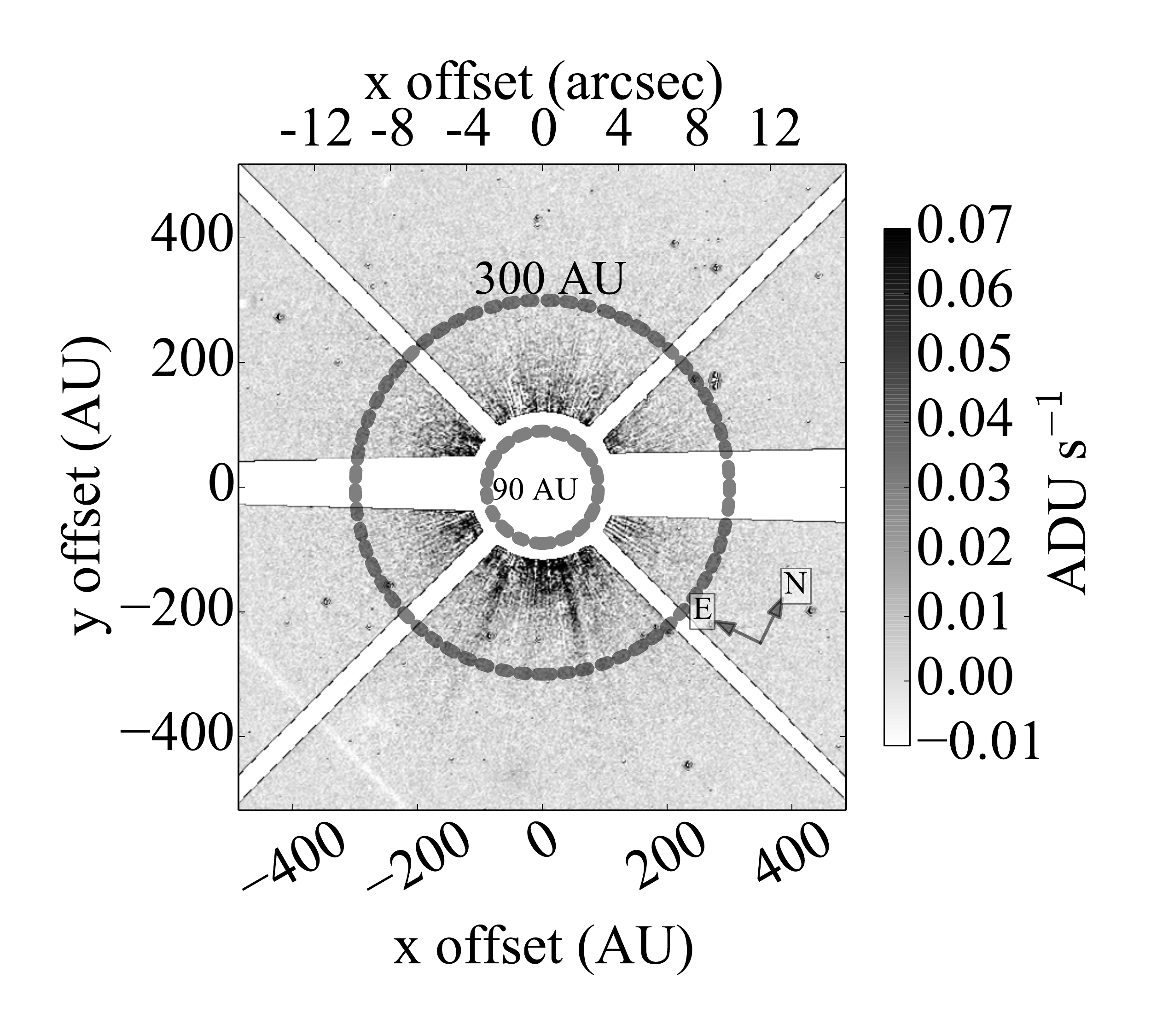} 
    \vspace{4ex}
  \end{minipage}
  \begin{minipage}[b]{0.5\linewidth}
    \centering
    \includegraphics[width=.65\linewidth]{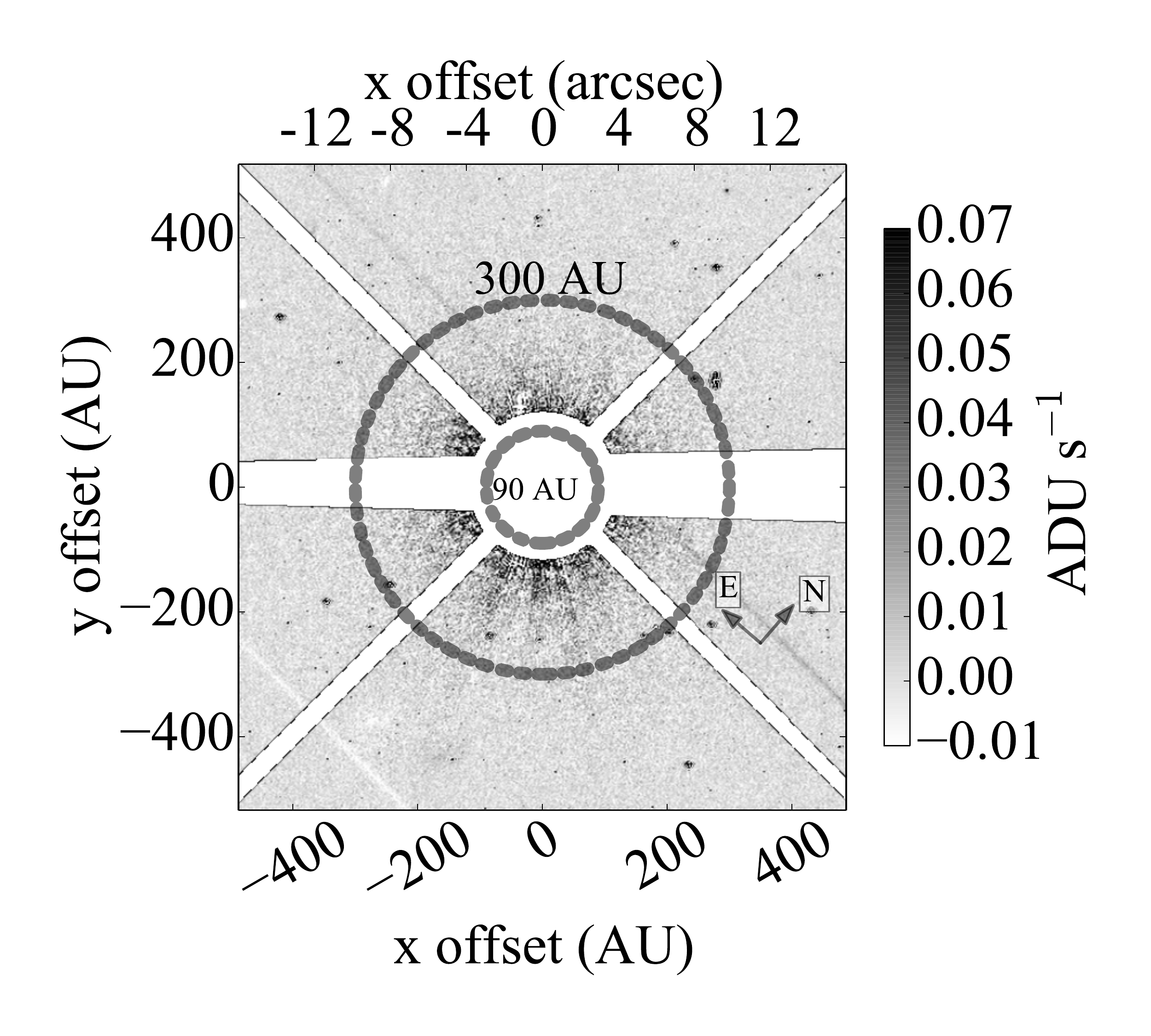} 
    \vspace{4ex}
  \end{minipage} 
  \begin{minipage}[b]{0.5\linewidth}
    \centering
    \includegraphics[width=.65\linewidth]{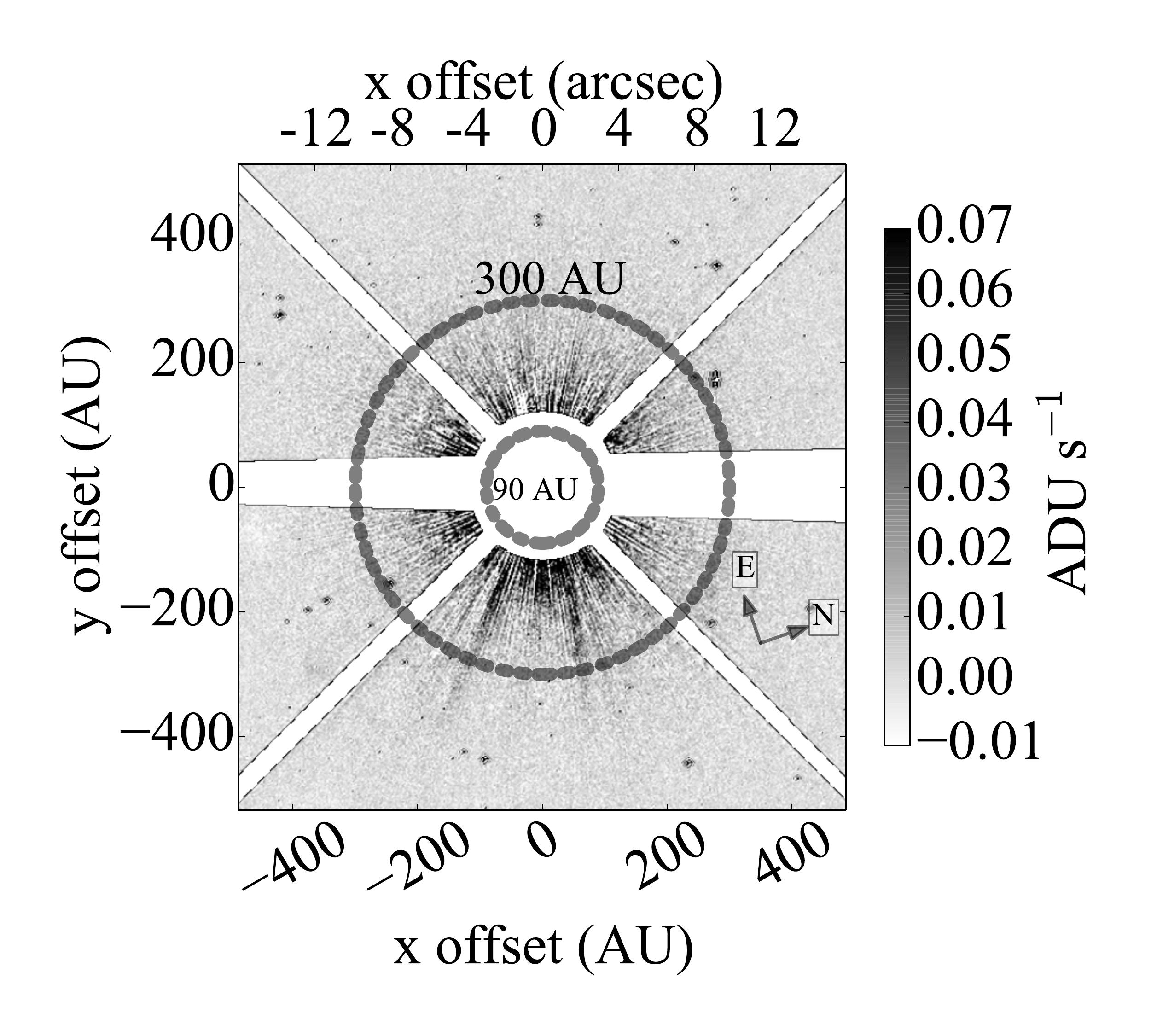} 
    \vspace{4ex}
  \end{minipage}
  \begin{minipage}[b]{0.5\linewidth}
    \centering
    \includegraphics[width=.65\linewidth]{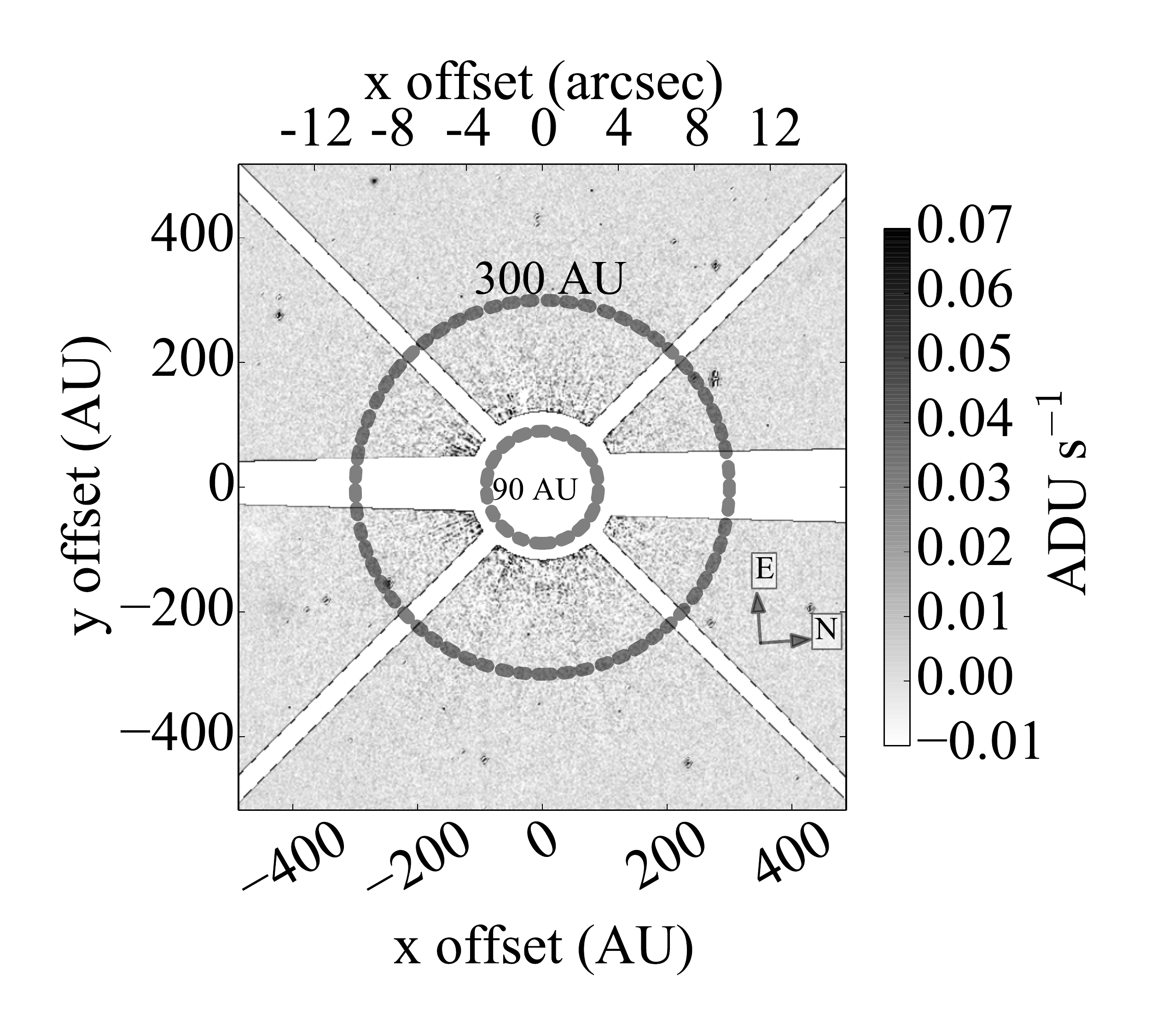} 
    \vspace{4ex}
  \end{minipage}
\caption{OSFi self-subtraction with (left column) and without (right column) the first image of each observing sequence. The first row is from epoch 1, the second is from epoch 2. Comparing the left column to the right column shows that the first image of each observing sequence contains additional low spatial frequencies, and so we exclude science 0 and science 3 from our set of reference images.}
\label{fig: science_0_3}
\end{figure*}

These results use only science images 1, 2, 4, and 5 from epochs 1 and 2. We found in self-subtraction that use of science image 0 or 3, corresponding to the first orbit in each sequence of images, shows a noisier residual than self-subtraction without either of these images, illustrated in Figure \ref{fig: science_0_3}. We may be seeing here that the first orbit in an \textit{HST} observing sequence is has lower pointing stability compared to subsequent orbits.

One potential problem with OSFi PSF subtraction is that it is optimized to find \textit{any} low spatial frequency residual. High spatial frequency normalization means that this residual could be entirely from any unnormalized low spatial frequencies in the science and/or reference PSF (i.e., from a difference in spectral type and/or focus between the two science images; see \S \ref{spec_defocus}). The absence of any low spatial frequency residual in Figure \ref{fig: self_subtract} shows that there is no detected instability or focus evolution effects between the two consecutive science images in either epoch.
\section{Detection Heuristics}
\label{heuristics}
We discuss a number of possible false-positive origins for the over-luminosity from OSFi reference-subtraction.
\subsection{Algorithm Effects}
\label{algo}
\subsubsection{Bootstrapping}
\label{boot}

We define bootstrapping as a form of PSF subtraction to measure algorithm throughput. This is done by adding a power law to a registered science image, im$_\text{sci}$, and then running OSFi PSF subtraction on that ``bootstrapped" image using the other regular science image in the same epoch as the reference image, im$_\text{ref}$. The residual should return the same PSF-subtracted radial profile as the original input power law as long as the added polynomial profile does not change the normalization coefficient, $\left(\frac{\sigma\left[ \mathscr{F}_H \left( \text{im}_\text{sci}\right) \right]}{\sigma\left[ \mathscr{F}_H \left(\text{im}_\text{ref} \right) \right]}\right)$. If the residual is not the same as the input, this would indicate a non-unity algorithm throughput, resulting in either over- or under-subtraction. If this effect is severe, it could amplify the noise in a non-detection to look like a detection.
\begin{figure*}[!p]
\centering
\plottwo{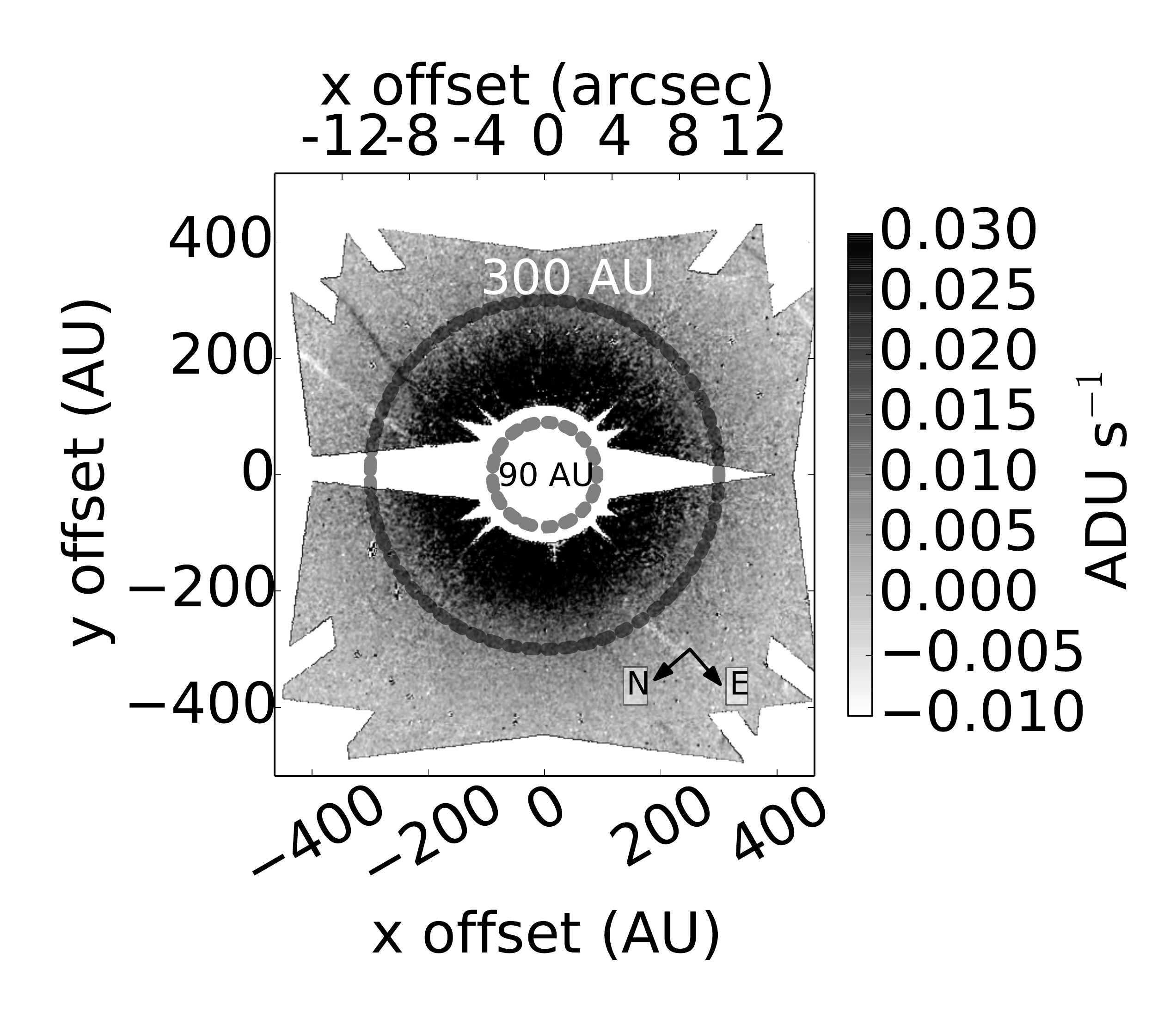}{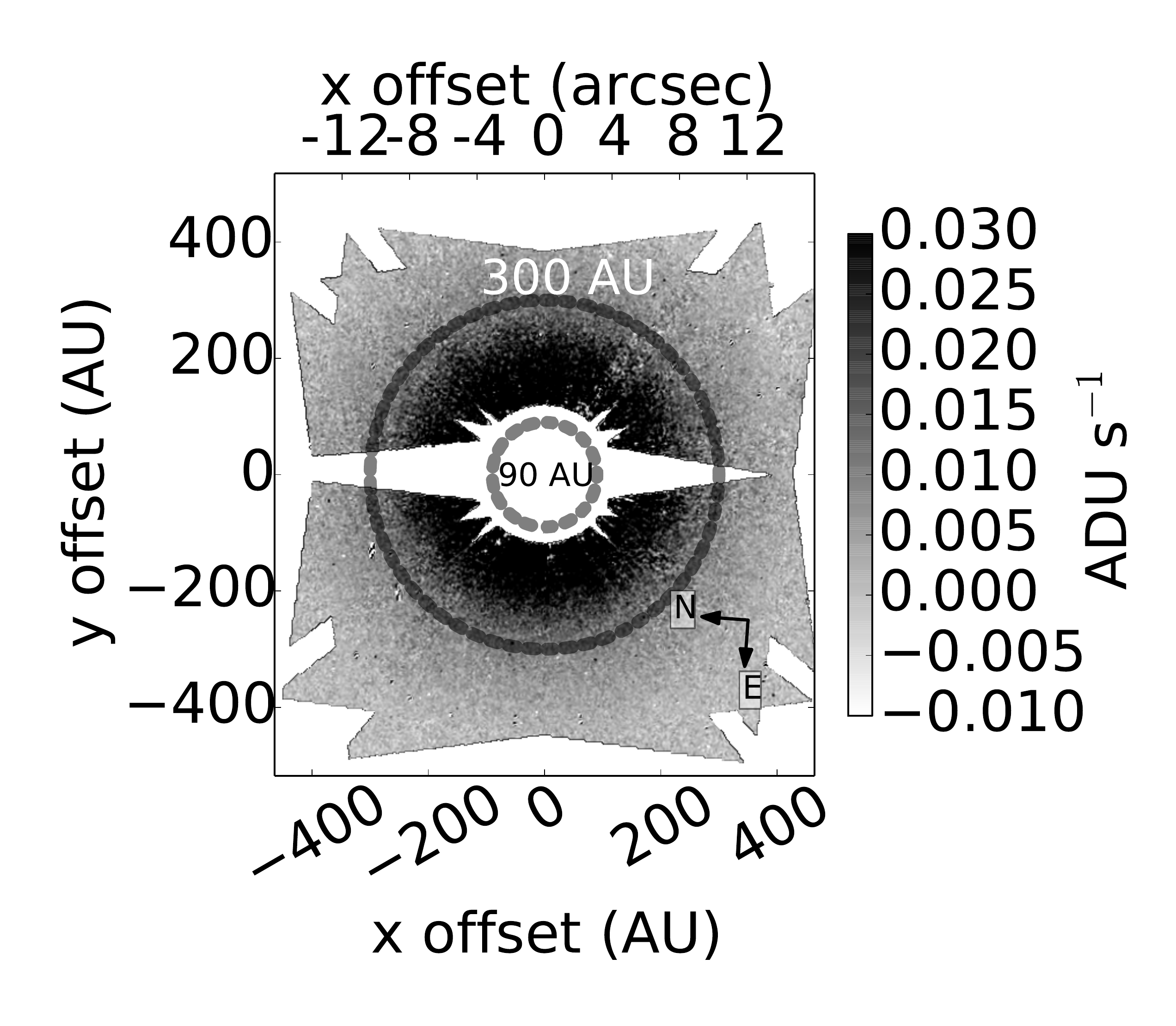}
\caption{Epoch 1 (left) and epoch 2 (right) bootstrapped OSFi PSF-subtracted images.}
\label{fig: bootstrap}
\end{figure*}

We chose to input the fitted polynomial profile from mean radial profile of the reference-subtracted images, but any other profile would serve the same purpose. We found that using a Gaussian 11 pixel standard deviation convolution kernel for $\mathscr{F}_L$ in equation \ref{eq: im_high} best recovers the added input radial profile after self-subtraction. No other filters we tried returned a similar unity throughput, including 5, 11, 21, and 31 pixel median kernels and 5, 21, and 31 pixel standard deviation Gaussian kernels. The two PSF-subtracted, bootstrapped images, are shown in Figure \ref{fig: bootstrap},

showing a similar residual halo component as with OSFI reference-subtraction (Figure \ref{fig: ref_subtract}). Figure \ref{fig: OSFi_rad} shows the OSFi bootstrapped radial profiles as the dashed lines, output from the input polynomial dashed-dotted line, illustrating that a Gaussian 11 pixel standard deviation convolution kernel yields close to unity OSFi algorithm throughput. We also confirmed that bootstrapping $(1/3)\times$(original polynomial profile) similarly recovers the expected output radial profile (not shown in Figure \ref{fig: OSFi_rad}).

The agreement between bootstrapping and reference-subtraction indicates that the recovered over-luminosity is not amplified or over-subtracted by our OSFi algorithm.

\subsubsection{OSFi vs. spider-normalized PSF Subtraction }
\label{spiders}
The classical iterative \textit{HST} PSF subtraction technique has been to remove the diffraction spiders by eye \cite[e.g., ][]{schneider}. This is similar to normalizing reference-subtraction to the cumulative flux ratio between the science and reference image in an aperture mask around the spiders, similar to the spider aperture mask used in our image registration pipeline (\S \ref{image_registration}). In this case, the coefficient in equation \ref{eq: algo} changes from $\left(\frac{\sigma\left[ \mathscr{F}_H \left( \text{im}_\text{sci}\right) \right]}{\sigma\left[ \mathscr{F}_H \left(\text{im}_\text{ref} \right) \right]}\right)$ to $\left(\frac{\Sigma\left(\text{im}_\text{sci}\right)}{\Sigma\left(\text{im}_\text{ref}\right)}\right)$, where $\Sigma$ refers to the cumulative flux within the spider aperture mask.
\begin{figure*}[!p]
\centering
\plottwo{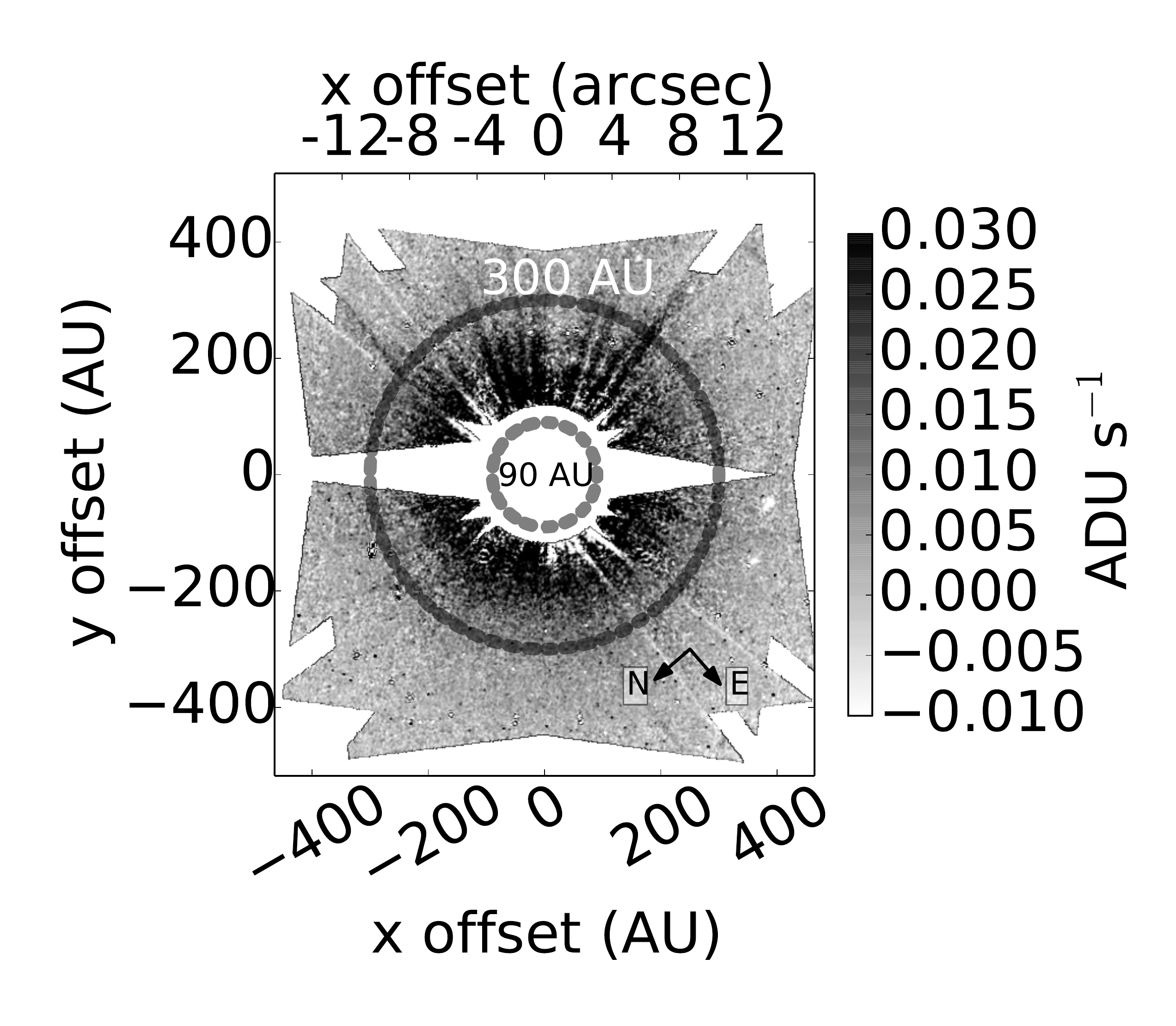}{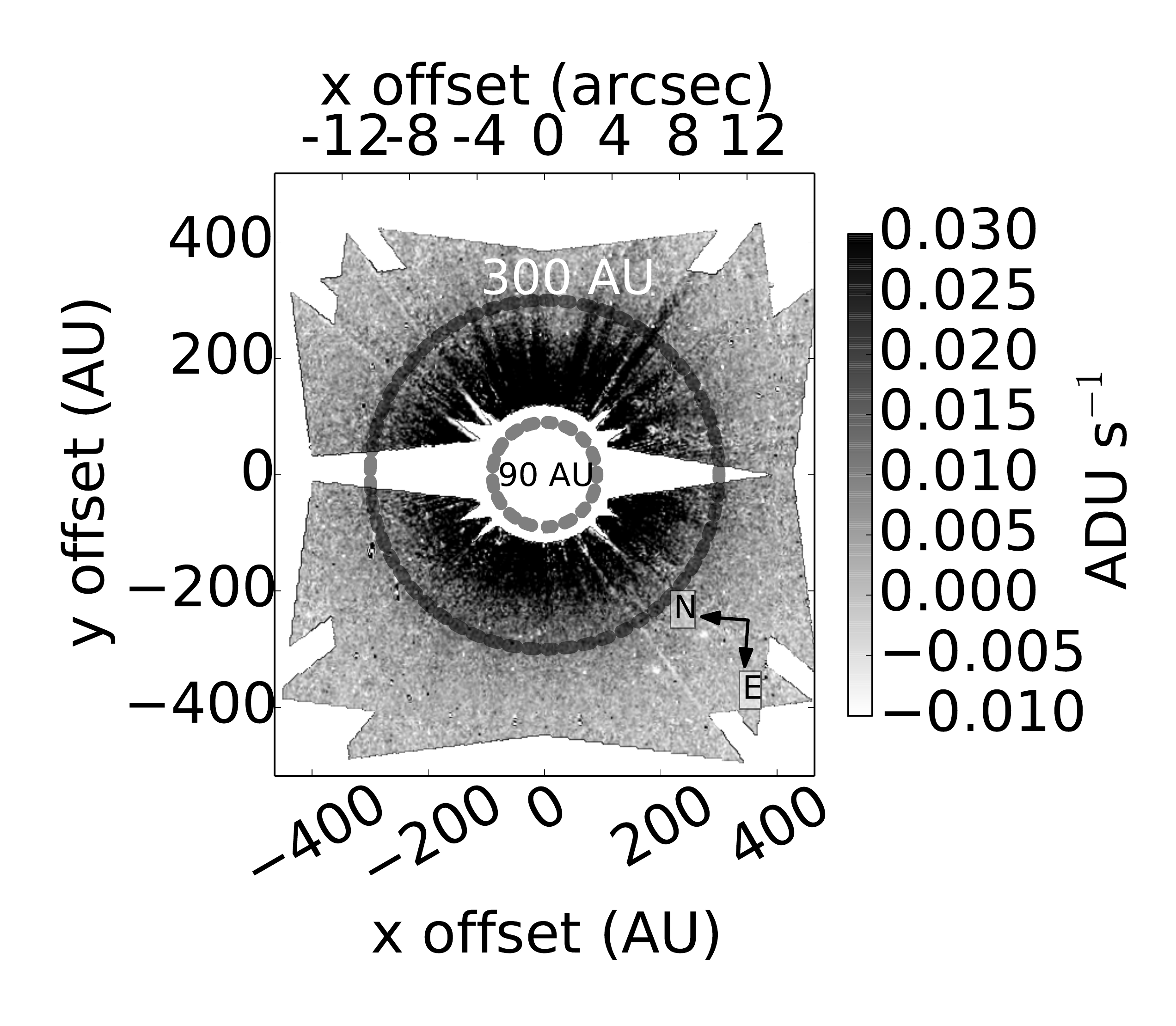}
\caption{Epoch 1 (left) and epoch 2 (right) reference-subtracted, spider-normalized images, where the reference star is HIP 117990 and the subtraction coefficient is now determined by normalizing reference and target images to the cumulative flux in four apertures around the spiders. This is a separate PSF subtraction method from OSFi normalization. }
\label{fig: spider_ref_subtract}
\end{figure*}
\begin{figure}[!h]
\centering
\plotone{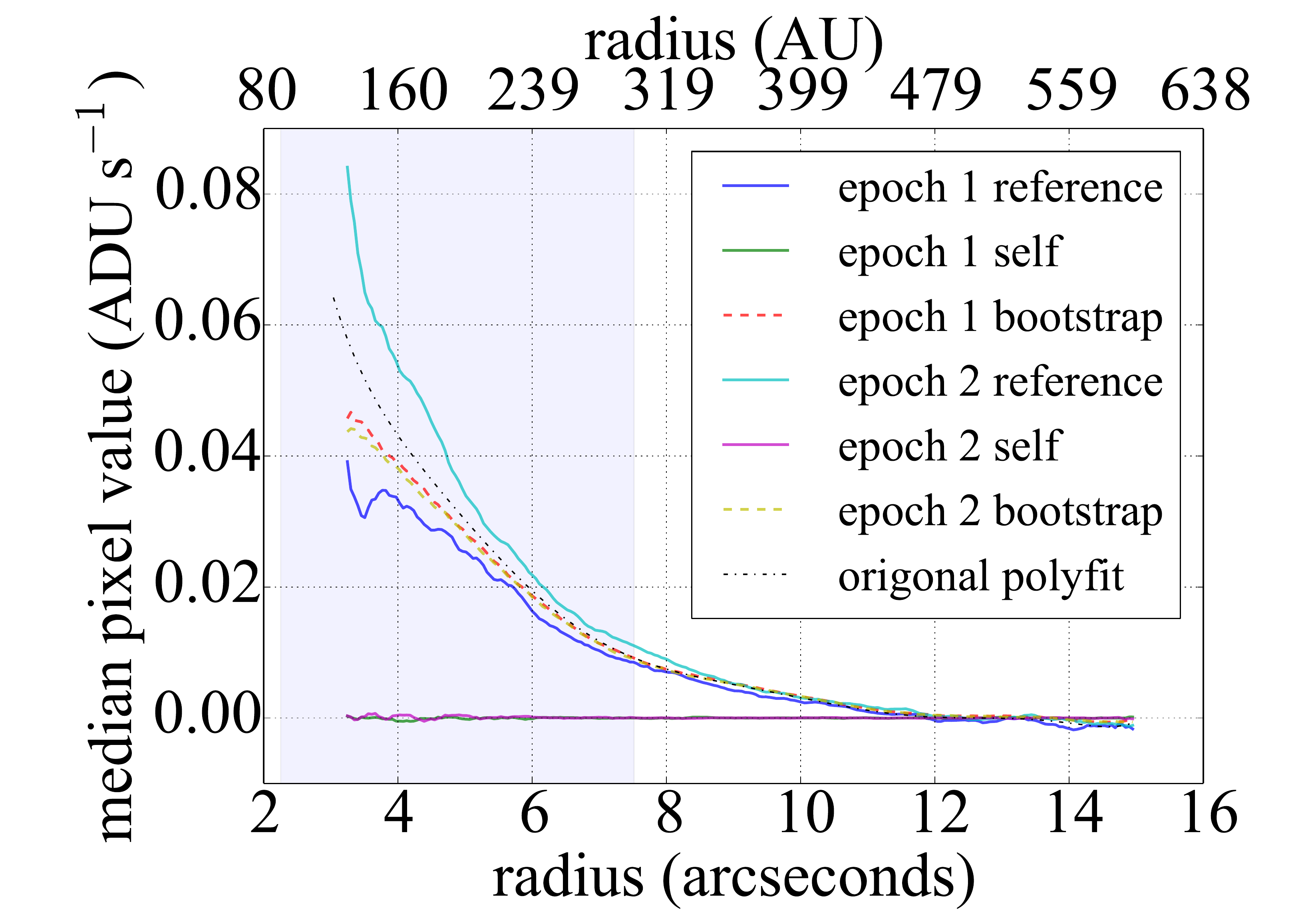}
\caption{Radial profiles of PSF-subtracted HR 8799 images using spider normalization, showing reference-subtraction, self-subtraction, and bootstrapping. The shaded 90--300 AU region represents the measured \su\; planetesimal belt.}
\label{fig: money_plot}
\end{figure}
\begin{figure*}[]
\centering
\plottwo{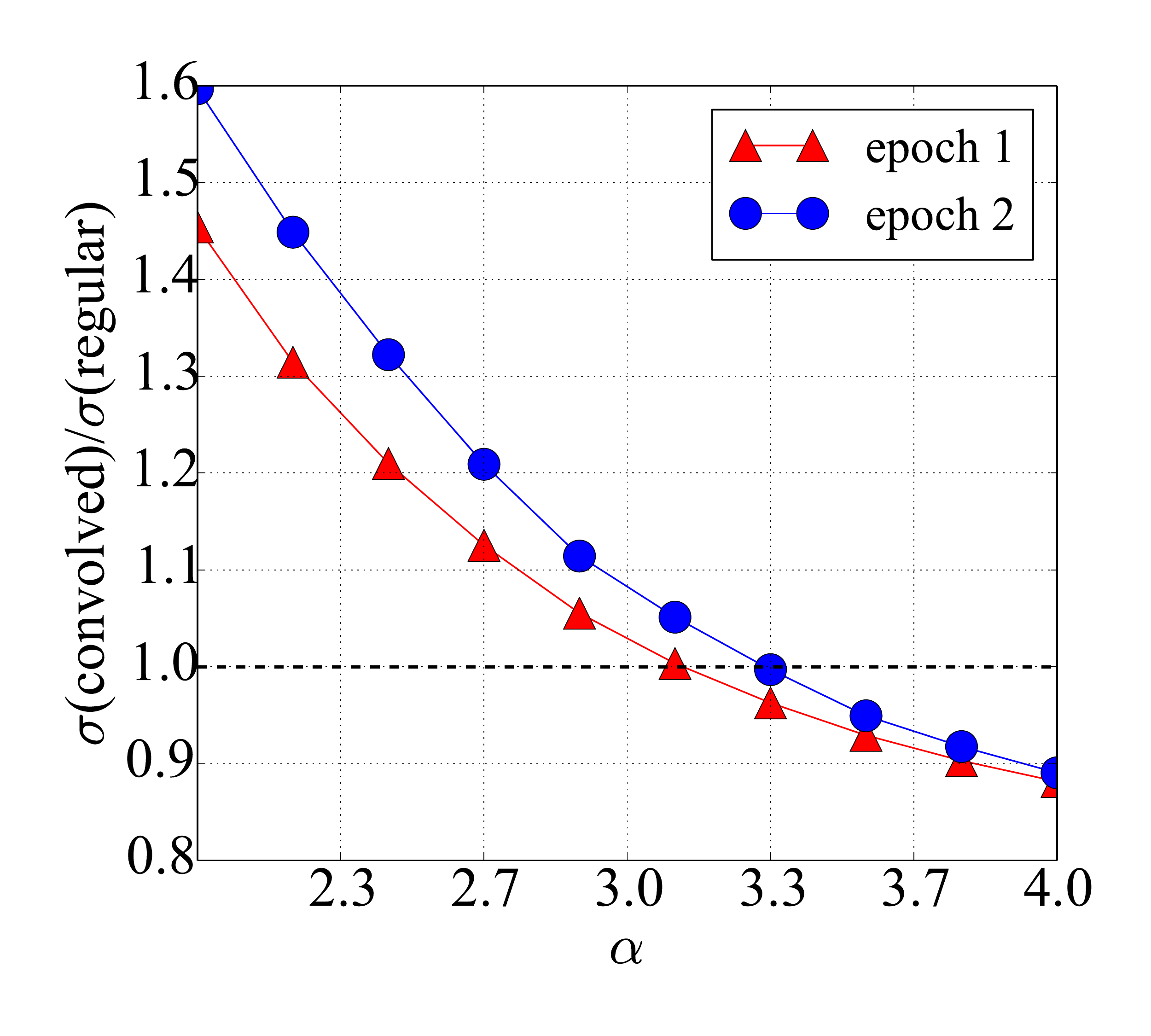}{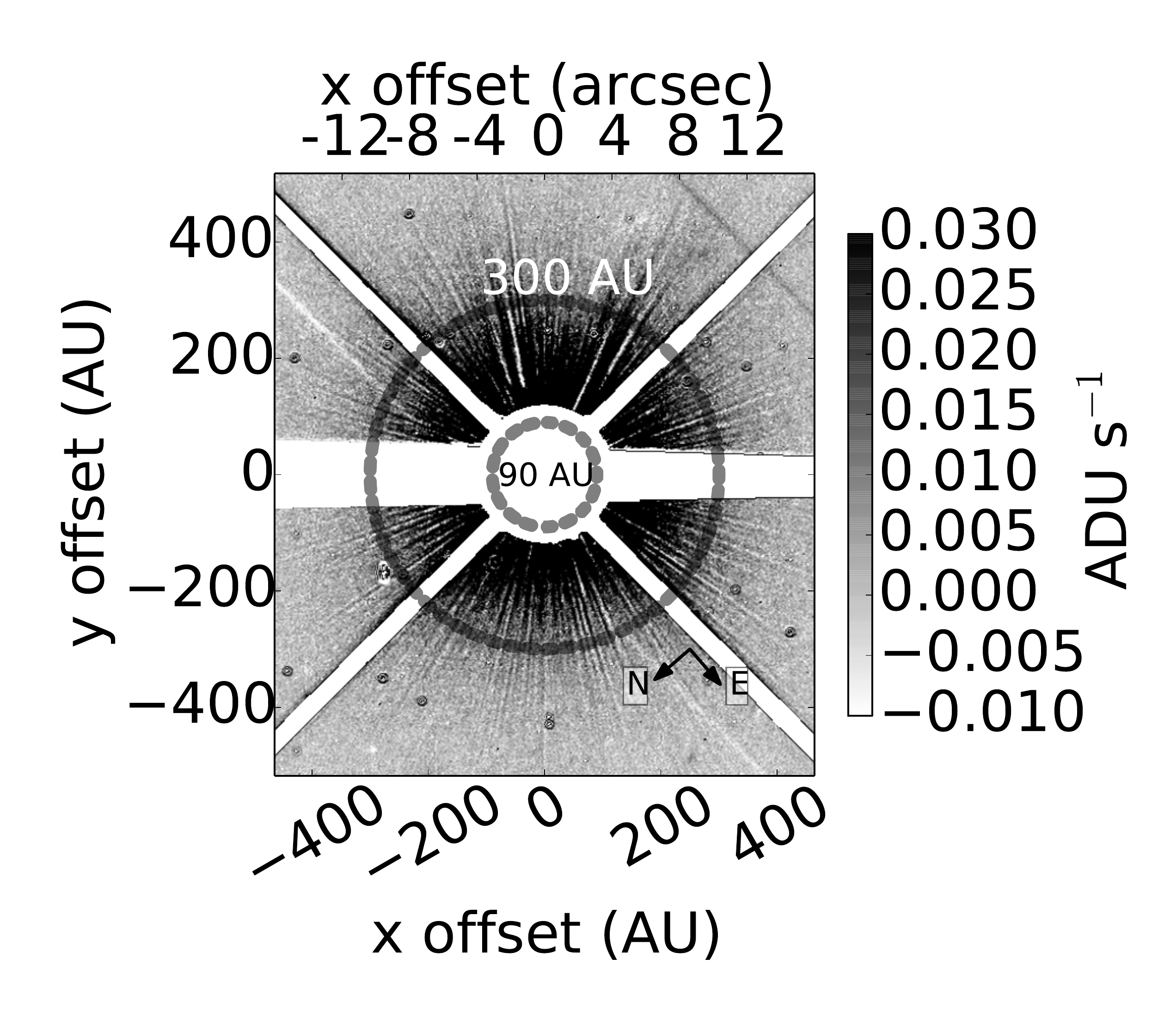}
\caption{Left: The robust standard deviation in $\mathscr{F}_H \left(\text{PSF}_\alpha\right)$---indicated as $\sigma$(convolved)---relative to regular OFSi reference-subtraction---$\mathscr{F}_H \left(\text{im}_\text{OSFi}\right)$, indicated as $\sigma$(regular)---vs. $\alpha$. $\alpha$ is the relative star-to-inner disk brightness. We use im$_\text{sci}=$ science image 2 and im$_{\text{sci}'}=$ science image 1. The black dotted line shows the desired amount of high spatial frequency noise in $\mathscr{F}_H \left(\text{PSF}_\alpha\right)$ needed to match the noise in $\mathscr{F}_H \left(\text{im}_\text{OSFi}\right)$. Right: PSF$_\alpha$ for $\alpha=3.2$, shown at the same pupil orientation as in Figure \ref{fig: ref_subtract}.}
\label{fig: conv}
\end{figure*}

Reference-subtracted images using spider-normalized PSF subtraction are shown Figure \ref{fig: spider_ref_subtract}. Their radial profiles are shown in Figure \ref{fig: money_plot} along with spider-normalized self-subtraction and bootstrapping results.

First, we see a general inconsistency between OSFi and spider-normalized reference-subtraction. These differences are most likely due to how the PSF subtraction algorithms determine the subtraction coefficient (e.g., different amounts of high spatial frequency leakage into the optimization region or spider apertures).

Second, in contrast to OSFi reference-subtraction, spider-normalized reference-subtraction shows an inconsistency between epochs 1 and 2. This discrepancy may be from the fact that the spiders are close to Nyquist sampled ($\sim$3 pixels across the FWHM), occasionally causing the brightest photon to fall undetected in between two pixels which may add a bias in calculating the normalization coefficient, $\left(\frac{\Sigma\left(\text{im}_\text{sci}\right)}{\Sigma\left(\text{im}_\text{ref}\right)}\right)$. Furthermore, radial profiles of spider-normalized bootstrapping (Figure \ref{fig: money_plot}) show a slight amount of over-subtraction. This is likely due to flux from the low spatial frequency disk (and/or instrumental halo) contaminating the spider flux in the bootstrapped image, thus causing a larger subtraction coefficient that results in over-subtraction. Although this bootstrapping over-subtraction is expected for spider normalization of close-to-face-on disks, an over-luminosity due to scattered light off of halo dust grains should in principle be consistent between different epochs and PSF subtraction algorithms after accounting for differences in throughput loss. This is not the case for spider normalization, suggesting that these inconsistencies may simply be from the non-optimal use of  spider-normalized PSF subtraction for close-to-face-on disks.
\subsection{PSF Effects}
\subsubsection{Inner Disk Convolution}
\label{inner_disk}
In Figures \ref{fig: ref_subtract} and \ref{fig: spider_ref_subtract}, shown at the same pupil orientation, high spatial frequency residual noise in the upper half of each image appears to be matching across both epochs, suggesting that these features are from the PSF. Thus, we consider the possible explanation that we are seeing the effect of the measured $\sim$6-15 AU inner disk (\su) on the STIS PSF.  This spatially unresolved inner disk should affect the PSF via convolution of a $\sim$6-15 AU annular kernel with the stellar PSF, ultimately causing some amount of high spatial frequency noise after subtracting the ``sharper,'' un-convolved reference PSF from the ``smoother," inner disk-convolved science PSF. However, the relative optical brightness of the inner disk to the stellar PSF component is unknown due to the unknown HR 8799 debris disk optical grain scattering properties, and so we set this as a free parameter in the equation below:
\begin{align}
\text{PSF}_\text{conv} &= \text{im}_\text{sci}(1-1/\alpha) + \left[\text{im}_\text{sci} \ast d_\text{inner} \right] (1/\alpha), \nonumber \\
\text{PSF}_\alpha &= \text{PSF}_\text{conv} - \text{im}_{\text{sci}'} \left(\frac{\sigma\left[  \mathscr{F}_H \left( \text{PSF}_\text{conv}\right) \right]}{\sigma \left[  \mathscr{F}_H \left( \text{im}_{\text{sci}'} \right) \right]}\right),
\label{eq: alpha}
\end{align}
where $\alpha$ represents the relative star-to-inner disk brightness and is an unknown scalar value to be optimized in the corresponding reference-subtraction, $d_\text{inner}$ is a 6-15 AU annular convolution kernel (3 to 7 pixels from the image center) whose sum is normalized to one, im$_\text{sci}$ is the input science image to be convolved with the $d_\text{inner}$ kernel, im$_{\text{sci}'}$ is the reference image to be used in self-subtraction (i.e., another science image in the same epoch as im$_\text{sci}$), and $\ast$ is the convolution operator. The two factors of $(1/\alpha)$ and $(1-1/\alpha)$ in equation \ref{eq: alpha} along with the cumulative normalization of $d_\text{inner}$ are set so that flux is conserved in transforming im$_\text{sci}$ into PSF$_\text{conv}$. We are running self-subtraction to observe only the PSF effects of inner disk convolution, whereas reference-subtraction would produce inner disk convolution effects in addition to the residual already seen in Figure \ref{fig: ref_subtract}. To find the optimal value of $\alpha$, we first measure the standard deviation of PSF$_\alpha$ in the same optimization region as in regular reference-subtraction after 11 pixel Gaussian high-pass filtering. We then compare this standard deviation to the standard deviation of the corresponding regular im$_\text{sci}$ OSFi reference-subtraction in the same high-pass filtered optimization region for a given value of $\alpha$. Standard deviation of the high-pass filtered PSF-subtracted image serves as a useful metric to measure the high spatial frequency noise coming from inner disk convolution.

Figure \ref{fig: conv} shows the relative standard deviation between PSF$_\alpha$ and regular OSFi reference PSF subtraction plotted vs. $\alpha$, indicating that the optimal $\alpha$ value between both epochs is $\alpha\sim3.2$. The right hand panel of Figure \ref{fig: conv} shows PSF$_{\alpha=3.2}$. Upon initial inspection, Figure \ref{fig: conv} shows that high spatial frequency noise is present in both positive and negative values. However, Figure \ref{fig: ref_subtract} only shows the equivalent positive ``spikes'' and is thus most likely not from the effect of imaging the inner disk. 
\subsubsection{Pointing Accuracy}
\label{pointing}
It could also be possible that the residual we are seeing is a result of a difference in alignment of the STIS coronagraph wedge (i.e., FPM) with respect to the star center between the science and reference images. The instrumental PSF component from the FPM is at the same position with respect to the image center for every image, but pointing accuracy can cause the star position to move with respect to the FPM. We can test this accuracy by measuring the difference in offsets of the reference and science images from image registration. The width of the FPM at the position covering the star is $\sim$45 pixels, and we find a mean radial offset difference between science and reference images of $\sim$0.07 pixels, or $\sim$3 miliarseconds, and so this small relative pointing difference difference with respect to the FPM size is an unlikely cause of the observed over-luminosity.
\subsection{Breathing, Spectral Difference}
\label{spec_defocus}

The origin of over-luminosity may be from defocus between the reference and science images. As mentioned in \S\ref{data}, the reference star was observed after the science images in epochs 1 and 2. Accordingly, the same pointing difference to slew from the science to the reference star can cause the same amount breathing defocus in both epochs, due to the same difference in incident sunlight angle with respect to \textit{HST} (J. Krist, private communication), which could explain the matching radial profiles between the two epochs.

In addition to focus, a difference in spectral type between the science and reference star can also cause a residual after PSF subtraction that looks like a close-to-face-on disk \citep{grady}. HR 8799 spectral type characterization varies between A5V and F0V with characteristics of an anomalous $\lambda$ Boo star \cite[]{gray}, and the reference star, HIP 117990, is an F2IV \cite[]{refstar}. Despite this difference, the broadband optical colors are closely matching: $(\text{B}-\text{V})_\text{HIP 117990}=0.26,\; (\text{B}-\text{V})_\text{HR 8799}=0.32$ \citep{color1}, and differing broadband spectral shape would be more deterministic of a residual halo feature than differing anomalous spectral lines.
\subsubsection{Tiny Tim Simulations}
To further investigate the possibility of a difference in focus and/or spectral type being the cause of the observed over-luminosity after PSF subtraction, we initially ran simulations with Tiny Tim, an \textit{HST} PSF simulator \cite[]{tinytim}. Using the closest matching \citet{pickles} synthetic spectra to the HR 8799 and reference star spectral type and the typical 5.8 $\mu$m defocus caused from breathing \cite{grady}, our initial results showed that neither spectral difference nor defocus are consistent with the observed over-luminosity. However, further evidence is needed to fully rule out instrumental PSF effects. Ultimately, to rule out the origin from spectral difference and/or coronagraph effects we would need real spectra for both the reference and science star over the full $0.2-1{\mu}$m STIS imaging bandpass used in a STIS PSF simulator that includes effects from the coronagraph. The STIS Tiny Tim PSF simulator does not include a coronagraph FPM or Lyot stop, which can cause low spatial frequency residuals from misalignment, and for this reason we instead ran OSFi PSF subtraction on another similar dataset, described below. 
\subsubsection{HD 10647}
If breathing defocus and/or spectral difference is the source of over-luminosity, we should see a similar radial profile after OSFi PSF subtraction of a different STIS target that is similar to HR 8799 in color difference and angular separation between the corresponding target and reference star. A similar color difference should create a similar effect over the STIS bandpass, and a similar angular separation should create a similar breathing defocus effect\footnote{The absolute angle of sunlight (i.e., position on the sky) will also matter in determining the amount of breathing defocus in addition to the relative target-reference separation, and so the separation alone should only give us a rough order of magnitude estimate. However, as we will see below, there are likely additional factors that cause breathing effects.}. 

Thus, we used the Canadian Astronomical Data Center (CADC) \textit{HST} Cache \citep{hst_cache}, astroquery \citep{astroquery}, and MAST to obtain STIS observations of HD 10647, a known edge on disk \citep{hd10647_mast}. The HD 10647 and HR 8799 datasets are closely matching in color difference and angular separation between the target and reference star used in each observing sequence: $\Delta(B-V)_\text{(HR 8799 - reference)}=0.06$ \citep{color1},  $\Delta(B-V)_\text{(HD 10647 - reference)}=0.05$ \citep{phot_cat}, $\Delta(\text{separation})_\text{(HR 8799 - reference)}$ $=$ $11.3^\circ$, and $\Delta(\text{separation})_\text{(HD 10647 - reference)}$ $=$ $9.1^\circ$ \citep{sep}. Since HD 10647 is an edge on disk, our rationale here is that if we see a radial profile similar to HR 8799 (i.e., resembling a close-to-face-on disk), we know in this case we are seeing instrumental and/or bandpass effects, which would suggest that this is also what we are seeing with HR 8799.

Initially, we chose to use the images with exposure times scaled to match the target (HD 10647) to reference (HD 7570) flux ratio, so that all images had the same cumulative detector flux, as with HR 8799\footnote{We name the consecutive HD 10647 observations, identified in the MAST dataset as OB1J09010, OB1J09020, OB1J10010, and OB1J10020 as science 0a, 1a, 2a, and 3a, respectively, and use the only scaled reference star observation, OB1J11030.}. We then applied the same OSFi pipeline as described above to this sequence, optimizing throughput via bootstrapping with a 21 pixel (standard deviation) Gaussian kernel (see Figure \ref{fig: hd10647} below). Each HD 10647 PSF-subtracted image was also flux-normalized based on the HD 10647 to HR 8799 flux ratio, which we determined using the spiders for each target.
\begin{figure}[!ht]
\centering
\plotone{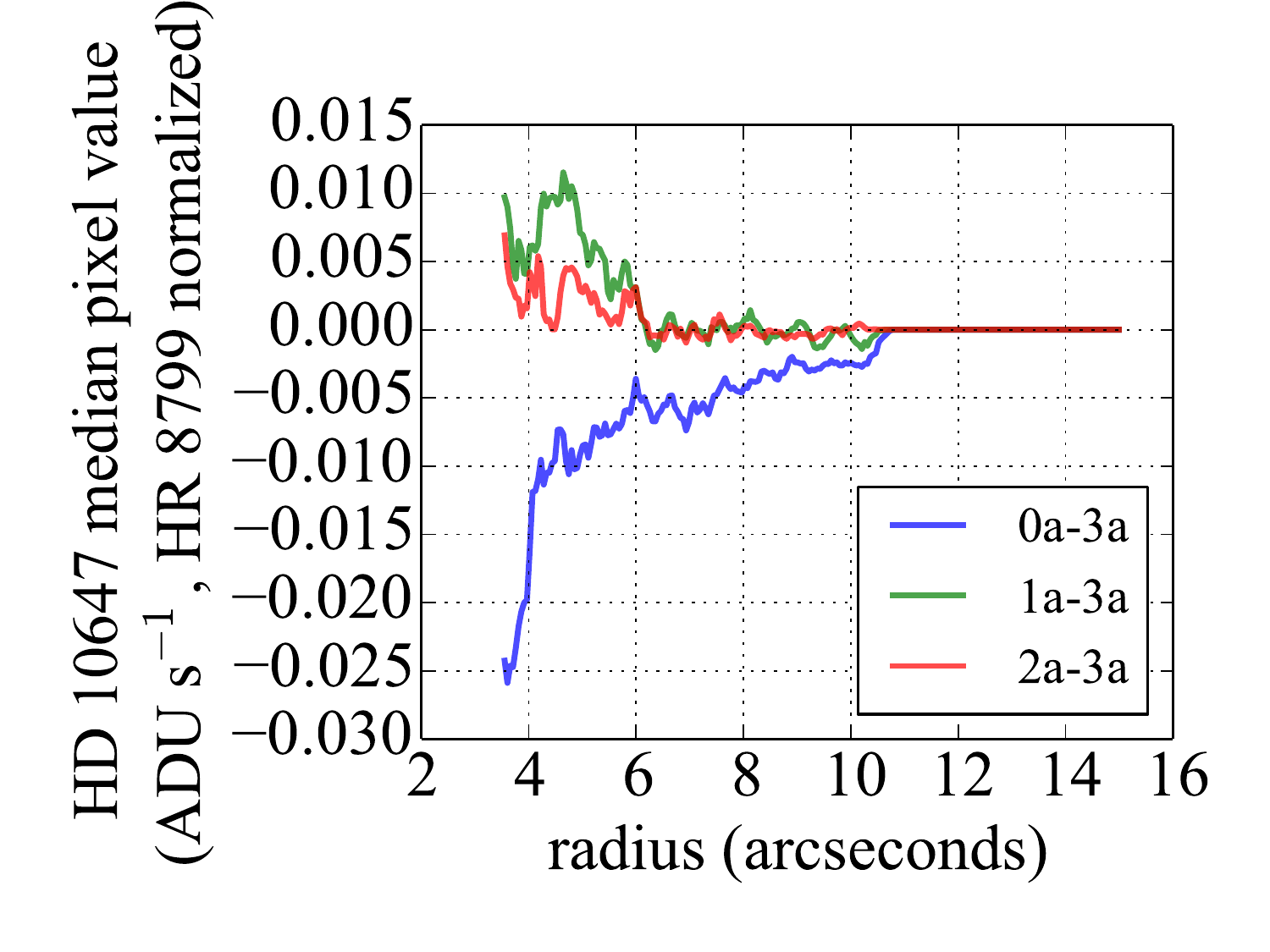}
\caption{OSFi self-subtracted radial profiles of HD 10647 \citep{hd10647_mast}, using science 3a as the reference image and science 0a (blue), 1a (green), and 2a (red) as the science images. We can see that 2a-3a has the most stable self-subtraction, indicating that there is a focus evolution throughout the HD 10647 observing sequence.}
\label{fig: focus_evolution}
\end{figure}
\begin{figure*}[!ht]
\centering
\plottwo{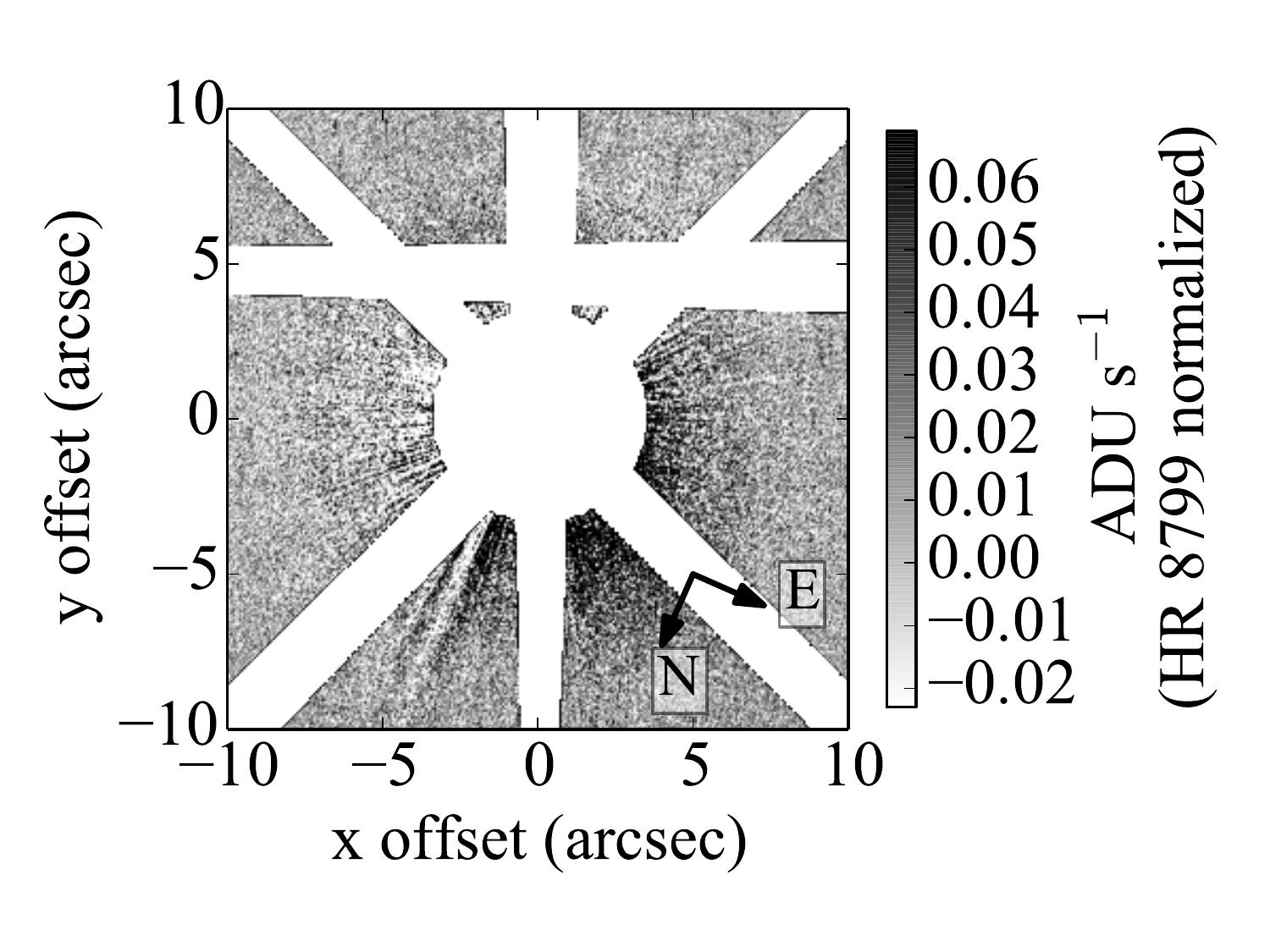}{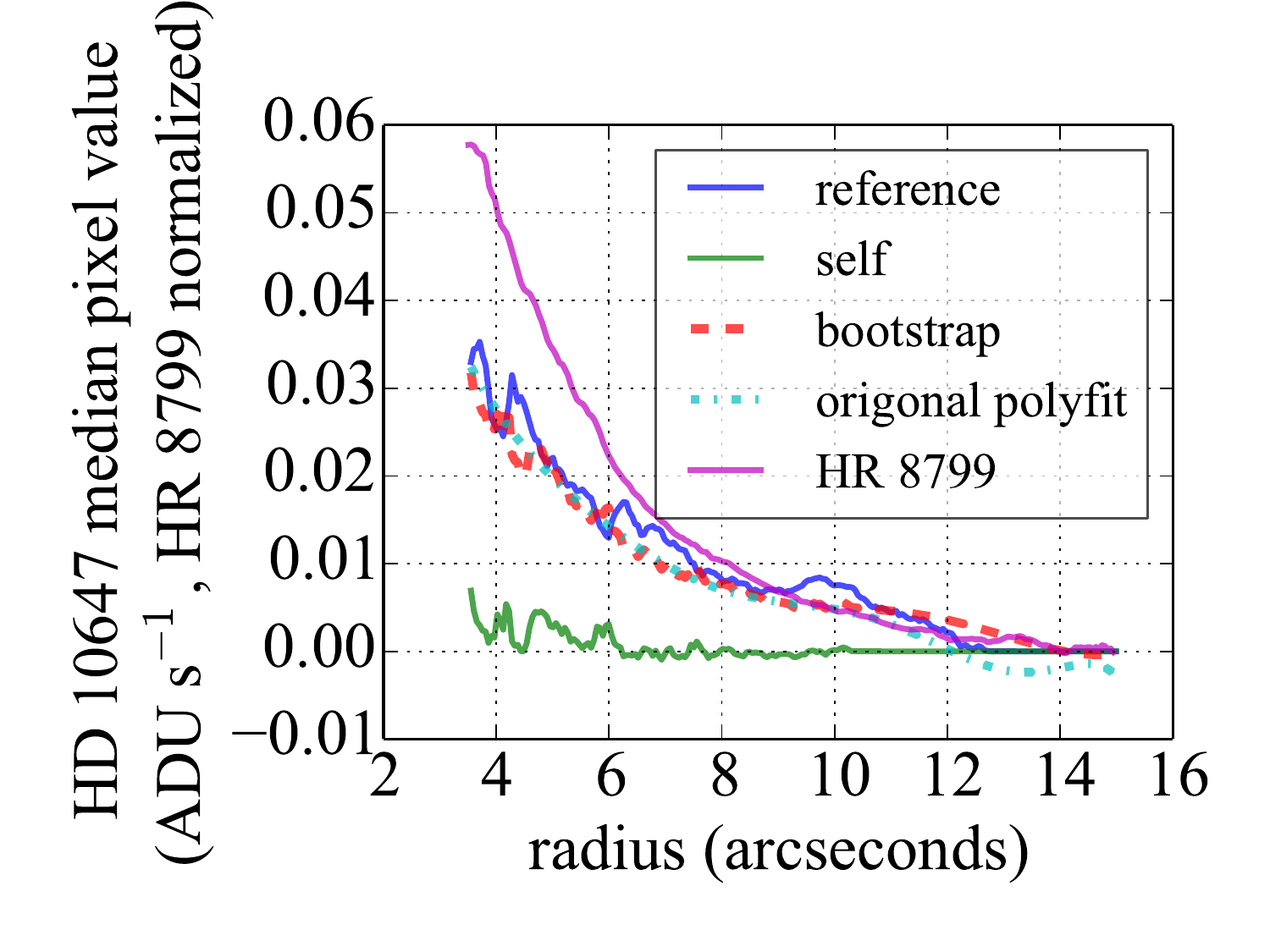}
\caption{Our OSFi reference-subtracted image of HD 10647 (left), a known edge on disk \citep{hd10647_mast}, and corresponding radial profiles (right). We do not detect the disk. We found the best match in bootstrapping (input: light blue, output: red) when using a 21 pixel (standard deviation) Gaussian kernel. To prevent effects from focus evolution, HD 10647 reference-subtraction is computed using 3a - reference and self-subtraction using 2a-3a. The similar HD 10647 and HR 8799 radial profiles (dark blue and pink, respectively) suggest that we are not seeing an astrophysical disk in either dataset.}
\label{fig: hd10647}
\end{figure*}

An initial run through of the OSFi pipeline on HD 10647 showed that the science images experience a focus evolution throughout the observing sequence. OSFi self-subtracted images on this sequence, using science 3a as a reference, are shown in Figure \ref{fig: focus_evolution}. We can see that the most stable self subtraction in this sequence is science 2a-3a, indicating that focus is changing throughout the sequence. We also see the same pattern using other images as the reference, further supporting the notion that two images are more stable with respect to one another if taken consecutively in the observing sequence. 

Using this information, we only selected science 3a as the science image (taken just before the reference image) in our OSFi reference-subtraction, discarding science 0a, 1a, and 2a. Accordingly, our OSFi reference-subtracted image and various radial profiles of HD 10647 are shown in Figure \ref{fig: hd10647}. We do not detect the known disk, instead seeing a similar halo profile as with HR 8799. We note that both self- and reference-subtraction are considerably noisier here than with HR 8799, as expected due to (1) a factor of $\sim$10 smaller in exposure times, (2) using only one target image and therefore removing the speckle suppression effects of ADI, and (3) the possible presence of additional focus evolution between science 3a and the HD 10647 reference star (we saw no such evidence for a similar focus evolution in HR 8799 self-subtraction). 

Figure \ref{fig: hd10647} shows that the reference-subtracted HD 10647 radial profile is within a factor of $\sim$2 of the HR 8799 reference-subtracted radial profile, suggesting that we are seeing a similar non-astrophysical source of over-luminosity in both targets. Thus, we proceed in modelling the HR 8799 debris disk from optical photometry assuming a non-detection, using the detected over-luminosity as the noise floor of our OSFi post-processing technique.
\section{Dust Disk Model, Upper Limits}
\label{science}
Assuming the observed over-luminosity is a non-detection, in the following argument we present an upper limit analysis on the mass in small (optical) grains, $M_\text{dust}$, in the HR 8799 debris disk system.  We assume the disk's reflecting dust particles are spherical with radius $a$ and cross section $\pi a^2 $. If so, an infinitesimal shell at distance $D_\text{shell}$ from the host star with grain density $\rho$ has a mass of
\begin{equation}
\label{eq: shell_mass}
M_\text{shell}\sim\frac{16\pi}{3} \rho \left(D_\text{shell}\right)^2 a \frac{L_\text{shell}}{L_\star},
\end{equation}
where $L_\text{shell}$ and $L_\star$ are the luminosity of the shell and star, respectively \citep{jura}. We assume $a=0.6\; \mu$m optical scattering grain radius (central wavelength in the 0.2-1 $\mu$m STIS bandpass) and small grain density of $\rho=2$ g cm$^{-3}$ as in \cite{rafael}. We determine $L_\star$ by integrating the Plank blackbody function over the STIS bandpass with $R_\star=1.34 R_\odot$ \citep{gray} and $T_\star=7250$ K \citep{sadakane}. $L_\text{shell}$ is the luminosity from a single unresolved shell of width one pixel within the measured $\sim$90-300 AU region. 

To determine $L_\text{shell}$ we first recompute the reference-subtracted images in Figures \ref{fig: ref_subtract} and \ref{fig: spider_ref_subtract} using the ``PHOTFLAM'' header parameter in the initial pipeline-reduced {\tt{.crj}} files to obtain OSFi and spider reference-subtracted images in physical units of 
$f_\lambda={\rm erg}~{\rm cm}^{-2}~{\rm sec}^{-1}~{\rm \AA}^{-1}$ \citep{stis_handbook}. This radial profile is then extrapolated via fifth order polynomial fitting, similar to \S\ref{algo}.1, beyond the 3.5 arcsecond saturated region to 90 AU (2.3 arcseconds) in order to cover the entire measured region of the planetesimal belt. The entire image is finally multiplied by 5 to obtain a ``5 $\sigma$'' upper limit\footnote{If the residual in the PSF-subtracted image is a non-detection, this represents a ``1 $\sigma$'' threshold, assuming Gaussian statistics, for which an astrophysical detection in that image should be at least 5 times brighter.}.

We then compute the cumulative flux, $f_\text{shell}$, in a one pixel wide annulus within the dust belt 90-300 AU region by taking the median value in each annulus times the number of pixels in the annulus, removing any additional bad pixels and instrumental high spatial frequencies. The corresponding shell luminosity at the distance of HR 8799, $d_\text{HR 8799}$, is
\begin{equation}
\label{eq: L_shell}
L_\text{shell}=f_\text{shell}\; 4 \pi \left(d_\text{HR 8799}\right)^2
\end{equation}
Combining equations \ref{eq: shell_mass} - \ref{eq: L_shell}, the dust belt mass is then computed as a discrete sum over one pixel-wide annuli within the \su\; planetesimal belt region:
\begin{equation}
\label{eq: total_mass}
M_\text{dust} \leq \frac{16\pi}{3}\; \frac{\rho}{L_\star}\; a\; \sum_\text{shell = 90 AU}^\text{300 AU} \left[5 L_\text{shell}\;\left(D_\text{shell}\right)^2\right]
\end{equation}

The resulting mass upper limits for OSFi normalization epoch 1, OSFi normalization epoch 2, spider normalization epoch 1, and spider normalization epoch 2 are $1.4\times10^{-4}\;M_\oplus,\; 1.2\times10^{-4}\;M_\oplus,\; 1.1\times10^{-4}\;M_\oplus,\; \text{and } 1.7\times10^{-4}\;M_\oplus$, respectively, yielding a final upper limit on the mass in small dust grains of
\begin{equation}
\label{eq: upper_limit}
M_\text{dust} \leq 1.7\times10^{-4}\; M_\oplus.
\end{equation}
\su\; (Table 3 and \S4.2) also determine the dust mass within HR 8799's 90-300 AU belt region, but from entirely different methods, using SED fitting from thermal emission at 24, 70, and 160 $\mu$m. They find a mass of $M_\text{dust} \sim 1.9\times10^{-2}~M_\oplus$ in grains between 1 and 10 $\mu$m in the 90 -- 300 AU planetesimal belt. 
They assume a standard collisional cascade grain size distribution power law of $n(a)\propto a^{-q}$ with $q=3.5$ \citep{col1}. Integrating over this size distribution from the minimum to maximum grain sizes for the infrared (1-10~$\mu$m) and optical (0.2-1~$\mu$m) observations\footnote{Here we are assuming the light that is scattered or re-emitted by dust grains is the same wavelength as the grain size.} gives the mass in each size bin
\begin{equation}
\label{eq: q_limit}
m(a) \propto {\int_{a_{\rm min}}^{a_{\rm max}}}n(a) \rho a^3 da \propto a^{0.5} \Big|_{a_{\rm min}}^{a_{\rm max}}. \\
\end{equation}
\su\; finds an optical grain dust mass consistent with their observations of $5\times10^{-3}~M_\oplus$, almost an order of magnitude larger in mass than the upper limit found above.  In other words, these \textit{HST} observations imply far less mass in small grains than one would expect from observations of dust at longer wavelengths and a collisional power-law size distribution.

The blowout limit for HR~8799, using $L_\star=4.9~L_\odot$, $M_\star=1.5~M_\odot$ \cite[]{gray} and $\rho=2~{\rm g~cm}^{-3}$ is $a_{bl}\sim2~\mu$m. 
As our observation wavelength of 0.2-1 $\mu$m is smaller than $a_{bl}$, a shallower-than-collisional-cascade power-law for optical grains makes sense (i.e., there should be less mass in grains smaller than $a_{bl}$ compared to what is expected from steady state collisions).  Thus, as expected, the optical grains we are sensitive to must be ejected from the HR~8799 system, removed faster than they can be produced in collisions. 

In order to estimate the mass-loss rate from the system by ejected small grains, we must estimate a timescale over which these grains are visible.  As these grains are on hyperbolic orbits, we will use the orbital timescale of barely bound grains to conservatively estimate the ejection timescale.  Barely bound grains will have highly eccentric orbits, with pericenters within the ``birth ring'' of parent bodies.  

Assuming a pericenter at the inner edge of the main planetesimal ring ($\sim$100~AU; \su), and an apocenter at the outermost observed extent of the halo \citep[$\sim$2000~AU;][]{Brenda}, the orbit of a barely bound grain will have a period on the order of ~$10^4$ years.  Assuming that the observed mass is ejected on this timescale, using equation \ref{eq: upper_limit} we find that $\pmb{\lesssim}$$1.7\times10^{-8}~M_\oplus$/year will be ejected from the system.

If this mass ejection rate is held constant over the age of HR 8799 \cite[$\sim$30~Myr;][]{marois10}, only $\pmb{\lesssim}$0.5~$M_\oplus$ should have been ground into dust and blown out of the system.  This order-of-magnitude upper limit on the mass is not worryingly large; it is smaller than estimates of parent body masses in most debris disk systems \cite[e.g.,][]{krivov}, including our own solar system \cite[e.g.,][]{solarsys}. Thus, this mass ejection rate alone does not imply a low-probability event such as a recent catastrophic collision to explain.

A few other stars with debris disks and extended dust haloes have been studied.  Due to its favorable edge-on viewing angle, $\beta$~Pic has been known for many years to host an extremely large dust halo visible in scattered light, extending at least $\sim$1000~AU from the star \citep[e.g.][]{kalas}.  Detailed collisional and radiative simulations have shown that the Vega system's halo is consistent with collisional production \citep{muller,sibthorpe}, and Fomalhaut is known to have an extended halo of very small grains \citep{espinoza}. \citet{su2015} suggest that the HD~95086 system's halo could be produced by collisions within a wide planetesimal belt, similar to the scattering disk component of the Kuiper Belt.

Interestingly, \textit{Herschel} and \textit{Spitzer} observations of HR~8799 find a large mass of dust grains in the outer halo \citep{Brenda}.  However, because this study is based on mid- and far-IR observations, it would not be sensitive to the very small dust grains seen in optical scattered light.  Grains observed by \textit{Herschel} and \textit{Spitzer} will be well above the blowout limit and thus bound to the star.  These larger grains would be ejected onto wide halo orbits by a different mechanism than the small grains discussed above.

The reader is reminded that in \S \ref{spec_defocus} we concluded that our observations show a deep non-detection of the HR 8799 debris disk. Thus, we conclude here that our analysis of a non-detection produces an upper limit on the mass loss rate of small dust grains in HR 8799 that is consistent with observations of other systems. %

\section{Summary \& Conclusions}
\label{conclusion}
We have developed a new PSF subtraction algorithm, called Optimized Spatially Filtered (OSFi) normalization, for direct imaging of close-to-face-on disks using space telescope reference PSF subtraction. The algorithm is optimized to normalize PSF subtraction to the high spatial frequency noise in each reference and science image, preventing contamination from a residual low spatial frequency close-to-face-on disk. Unlike  the classical spider normalization reference PSF subtraction technique, OSFi normalization is not affected by throughput contamination from a close-to-face-on disk.

After applying this algorithm to \textit{HST}/STIS HR 8799 data, our main findings are as follows:
\begin{itemize}
\item A low spatial frequency residual is present after reference-subtraction and is consistent between the two epochs separated by a year.
\item In comparison, self-subtraction between two separate roll angles within a single epoch yields no residual, showing that there is no detected focus evolution between science images.
\item Residual high spatial frequency noise is also present after reference PSF subtraction. We evaluate the possibility that this may be from imaging the PSF of the known inner $\sim$6-15 AU disk \citep{su, Brenda}, ultimately showing that this is unlikely and that we are likely seeing an instrumental effect.
\item Spider normalization reference-subtraction is inconsistent between the two epochs and also inconsistent with OSFi normalization, illustrating that classical spider normalization is unoptimized for face-on disks.
\item By running OSFi PSF subtraction on HD 10647, a similar dataset to the HR 8799 sequence, we demonstrate that the observed over-luminosity is likely from a combination of defocus and difference in spectral type over the STIS broadband between the reference and science images in both epochs.
\item Using the photometry of this result, we determine an upper limit on the planetesimal belt mass in optical grains, which is smaller than expected from the \citet{su} results, which assume a collisional cascade power law size distribution. This suggests that these optical grains are not steady state, as is expected for dust grains smaller than the blowout limit.
\item For these sub-blowout limit dust grains, we estimate an ejection timescale that, if constant over the lifetime of HR 8799, would grind down $\pmb{\le}0.5 M_\oplus$ in parent bodies, consistent with estimates for other known debris disk systems.
\end{itemize}

We gratefully acknowledge research support of the Natural Sciences and Engineering Council (NSERC) of Canada. SML gratefully acknowledges support from the NRC Canada Plaskett Fellowship. This research used the facilities of the Canadian Astronomy Data Centre operated by the National Research Council of Canada with the support of the Canadian Space Agency. STScI is operated by the Association of Universities for Research in Astronomy, Inc., under NASA contract NAS5-26555. Support for MAST for non-\textit{HST} data is provided by the NASA Office of Space Science via grant NNX09AF08G and by other grants and contracts. The authors thank the anonymous referee for his or her comments and suggestions that have improved this manuscript.
\bibliography{bibtex.bib}

\end{document}